%% file: main.tex
\documentclass[journal]{IEEEtran}
\usepackage{amsmath,amsfonts}
\usepackage{algorithmic}
\usepackage{acronym}
\usepackage{array}
\usepackage{booktabs}
\usepackage{cite}
\usepackage{caption}
\usepackage{subcaption}
\usepackage{textcomp}
\usepackage{stfloats}
\usepackage{url}
\usepackage{siunitx}
\usepackage{verbatim}
\usepackage{graphicx}
\usepackage{subfiles}
\usepackage{color}
\def\BibTeX{{\rm B\kern-.05em{\sc i\kern-.025em b}\kern-.08em
    T\kern-.1667em\lower.7ex\hbox{E}\kern-.125emX}}
\usepackage{balance}
\usepackage{hyperref}
\graphicspath{{images/}}

\begin{document}

\include{utils/com}
\include{utils/comm}

\title{Multi-Band Wi-Fi Neural Dynamic Fusion}
\author{Sorachi Kato, Pu (Perry) Wang, Toshiaki Koike-Akino, Takuya Fujihashi, Hassan Mansour, Petros Boufounos
\thanks{Part of this paper was presented in ICASSP 2024 \cite{KatoWang24}.}
\thanks{The work of S. Kato was done during his visit and internship at MERL. He was also supported by Japan Society for the Promotion of Science~(JSPS) KAKENHI under Grant 23KJ1499.}
\thanks{PW, TK, HM, and PB are with Mitsubishi Electric Research Laboratories (MERL), Cambridge, MA 02139, USA.}
\thanks{SK and TF are with the Graduate School of Information Science and Technology, Osaka University, Suita, Osaka, Japan. }
\thanks{*Corresponding author: pwang@merl.com}
}

\maketitle

\begin{abstract}
Wi-Fi channel measurements across different bands, e.g., sub-$7$-GHz and $60$-GHz bands, are asynchronous due to the uncoordinated nature of distinct standards protocols, e.g., $802.11$ac/ax/be and $802.11$ad/ay. Multi-band Wi-Fi fusion has been considered before on a \emph{frame-to-frame} basis for simple classification tasks, which does not require fine-time-scale alignment. In contrast, this paper considers asynchronous \emph{sequence-to-sequence} fusion between sub-$7$-GHz channel state information (CSI) and $60$-GHz beam signal-to-noise-ratio~(SNR)s for more challenging tasks such as continuous coordinate estimation.
To handle the timing disparity between asynchronous multi-band Wi-Fi channel measurements, this paper proposes a multi-band neural dynamic fusion (NDF) framework. This framework uses separate encoders to embed the multi-band Wi-Fi measurement sequences to separate initial latent conditions. Using a continuous-time ordinary differential equation (ODE) modeling, these initial latent conditions are propagated to respective latent states of the multi-band channel measurements at the same time instances for a latent alignment and a {post-ODE} fusion, and at their original time instances for measurement reconstruction. We derive a customized loss function based on the variational evidence lower bound (ELBO) that balances between the multi-band measurement reconstruction and continuous coordinate estimation. We evaluate the NDF framework using an in-house multi-band Wi-Fi testbed and demonstrate substantial performance improvements over a comprehensive list of single-band and multi-band baseline methods. 
\end{abstract}


\section{Introduction}
\label{sec:intro}
\IEEEPARstart{W}{i-Fi} sensing, e.g., device localization and device-free human sensing, has received a great deal of attention in the past decade from both academia and industry. This trend has been manifested by the establishment of  802.11bf WLAN Sensing task group in 2020 to go beyond data transmission and meet industry demands for wireless sensing \cite{MeneghelloChen23, BlandinoRopitault23, ChenSong23, DuXie22}. 

Existing Wi-Fi sensing mainly relies on coarse-grained received signal strength indicator (RSSI) and fine-grained channel state information (CSI) at sub-$7$-GHz bands~\cite{YoussefAgrawala08, HoangYuen19, chen2017confi, DingWang19}. 
At a high frame rate, CSI reflects intrinsic channel statistics in the form of channel frequency responses (CFR) over subcarrier frequencies (delay) and multiple transmitter-receiver antenna pairs (angle). At the same time, it may experience channel instability due to even small-scale environment changes. 
On the other hand, mid-grained mmWave beam training measurements at $60$ GHz, e.g., beam {SNR}, have shown better channel stability over time \cite{KoikeWang20, GarciaLacruz20, YuWang22, BlancoMateo22, RubioWang23, RubioWang24}. These beam SNR measurements originate from sector-level directional beam training, a mandatory step for mmWave Wi-Fi to compensate for large path loss and establish the link between access points (APs) and users.
However, they suffer from low frame rates and irregular sample intervals due to the beam training overhead and follow-up association steps.

\begin{figure}
    \centering
    \includegraphics[width=1\hsize]{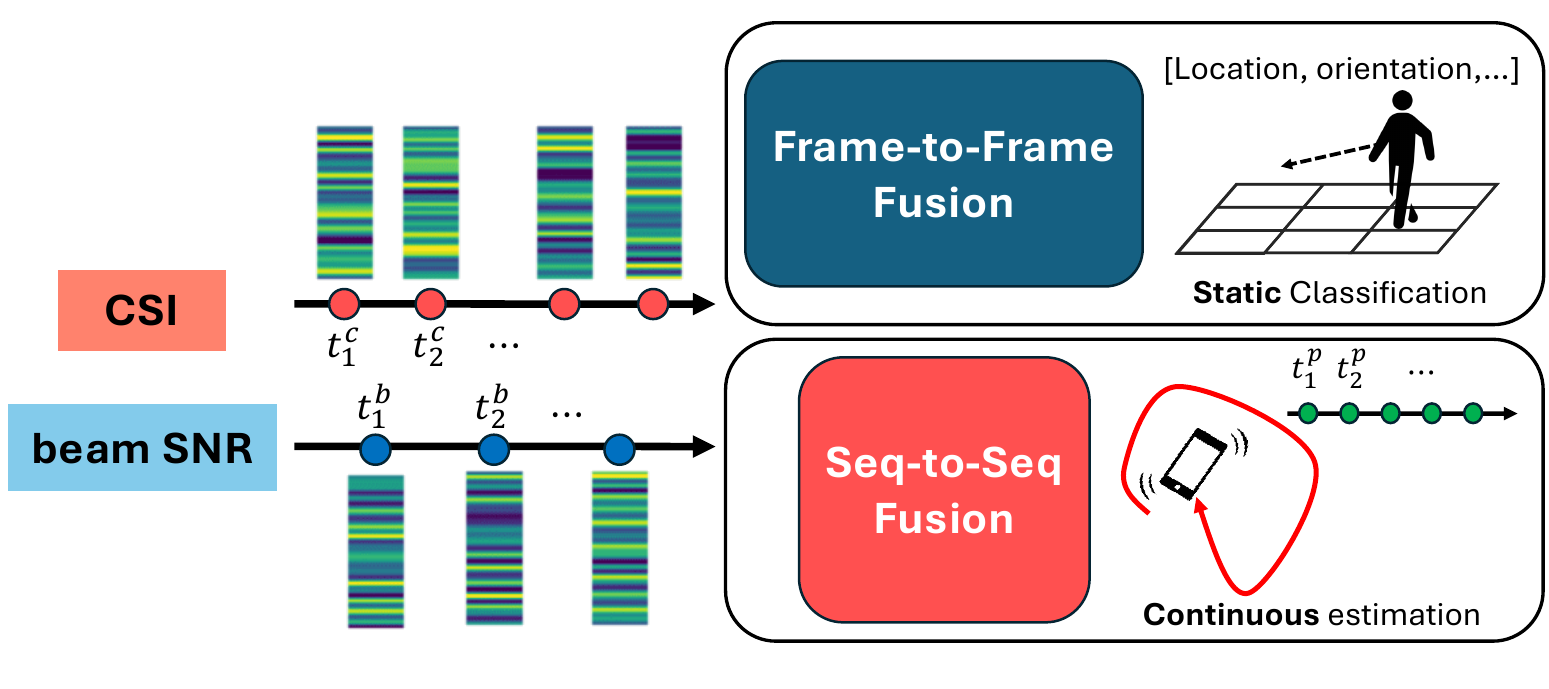}
    \caption{Multi-band Wi-Fi fusion from \emph{frame-to-frame} basis of \cite{YuWang22} for classification (top) to asynchronous \emph{sequence-to-sequence} basis for continuous-time regression (bottom).}
    \label{fig:frame_vs_sequence}
    \vspace{-0.1in}
\end{figure}

Fusion-based approaches have been considered in the literature for robustness and higher accuracy. Heterogeneous sensor fusion was studied between Wi-Fi and other modalities, e.g., Wi-Fi and vision~\cite{deng2023gaitfi}, Wi-Fi and ultra-wideband (UWB) \cite{YuWang22}.
Within Wi-Fi channel measurements, CSI and {RSSI} can be simply concatenated for joint feature extraction~\cite{AhmedArablouei19}.
It is also possible to fuse the phase and amplitude of the fine-grained {CSI} for localization \cite{DangSi19}.  
For multi-band Wi-Fi fusion, our previous work in \cite{YuWang22} appears to be the only effort considering the fusion between CSI at sub-$7$-GHz and beam SNR at $60$ GHz.
However, \cite{YuWang22} only considered simple classification tasks, e.g. pose classification (over $8$ stationary poses), seat occupancy detection ($8$ stationary patterns), and fixed-grid localization.
Despite being sampled at different time instances, both channel measurements can be simply combined on a \emph{frame-to-frame} basis as these asynchronous samples correspond to the same stationary label (e.g., pose, occupancy, location) and their respective sampling time becomes irrelevant for the fusion task; see the top plot of Fig.~\ref{fig:frame_vs_sequence} and notice the asynchronous time instances $t^c_n$ and $t^b_n$, $n=1, 2, \dots$.

For more challenging tasks of continuous-time object trajectory estimation using asynchronous multi-band Wi-Fi measurements, several challenges need to be addressed. First, asynchronous channel measurements at different bands need to be aligned to estimate object trajectory. As illustrated in Fig.~\ref{fig:frame_vs_sequence} (the bottom plot), there exists time disparity between the CSI measurements at $t^c_n$ and beam SNR measurements at $t^b_n$. In addition, there may exist time disparity between the input measurements at either $t^c_n$ or $t^b_n$, and the desired trajectory estimates at $t^p_n$. Second, mmWave beam SNRs are sampled at a much lower frame rate than the CSI measurements. In an ideal scenario, beam SNRs can be obtained at a frame rate of $10$ Hz for a typical beacon interval of $100$ ms. However, multiple users need to contend the channel time for (uplink) beam training during each beacon interval, resulting in a lower frame rate. Third, the contention-based channel access further results in irregularly sampled beam SNR measurements at AP for a given user.

To address the aforementioned challenges, we propose a multi-band Wi-Fi neural dynamic fusion (NDF) framework. This framework evolves from the stationary frame-to-frame basis in \cite{YuWang22}  (illustrated in the upper plot of Fig.~\ref{fig:frame_vs_sequence}) to a dynamic asynchronous \emph{sequence-to-sequence} basis (the bottom plot of Fig.~\ref{fig:frame_vs_sequence}), thus supporting more challenging downstream tasks, e.g., regression in a continuous space and continuous-time object trajectory estimation. 
The proposed multi-band NDF substantially extends our previous work on a beam SNR-only framework of \cite{RubioWang23} and \cite{RubioWang24} to a neural network architecture comprising multiple encoders, latent dynamic learning modules, a post-ODE fusion module, and multiple decoders, as depicted in Fig.~\ref{fig:overview}. 
Our main contributions are summarized below:
\begin{enumerate}
    \item To the best of our knowledge, this is the first effort to address multi-band asynchronous fusion between sub-$7$-GHz {CSI} and $60$-GHz beam SNR for trajectory estimation of moving objects. 
     \item We present a multi-encoder, multi-decoder NDF network in Fig.~\ref{fig:overview}. It utilizes the two encoders acting like an initial latent condition estimator for the two distinct input sequences, employs an ordinary differential equation (ODE) modeling \cite{chen2018neural, rubanova2019latent} for latent dynamic learning and latent state alignment, and fuses these aligned latent states via the post-ODE fusion module. 
      \item We consider multiple fusion schemes such as a multilayer perceptron (MLP) fusion, a pairwise interaction fusion, and a weighted importance fusion for the post-ODE fusion module. 
    \item We derive a loss function building upon the variational evidence lower bound (ELBO) between prior and approximate posterior distributions of the initial latent conditions as well as the likelihood of multiple decoder outputs. This ELBO-based loss function incorporates both unsupervised multi-band reconstruction loss and supervised coordinate estimation loss. 
    \item We build an automated data collection platform utilizing commercial-of-the-shelf 802.11ac/ad-compliant Wi-Fi routers and a TurtleBot as a mobile user. This platform continuously gathers CSI at $5$ GHz and beam SNR at $60$ GHz from the TurtleBot, while simultaneously recording its ground truth positions.
    \item We conduct a comprehensive ablation study on trajectory estimation performance, generalization capability, and interpretation using real-world experimental data. 
\end{enumerate}

The remainder of this paper is organized as follows.
Section~\ref{sec:problem_formulation} introduces the problem formulation, followed by a brief review of existing multi-band Wi-Fi fusion solutions. Section~\ref{sec:ndf} details the proposed multi-band NDF framework, with subsections dedicated to each module and the derivation of the loss function. Section~\ref{sec:evaluation} describes our in-house multi-band Wi-Fi data collection testbed and performance evaluation, followed by the conclusion in Section~\ref{sec:conclusion}.

\section{Problem Formulation and Existing Solutions}
\label{sec:problem_formulation}

\subsection{Problem Formulation}
\begin{figure*}[t]
    \centering
    \includegraphics[width=1\textwidth]{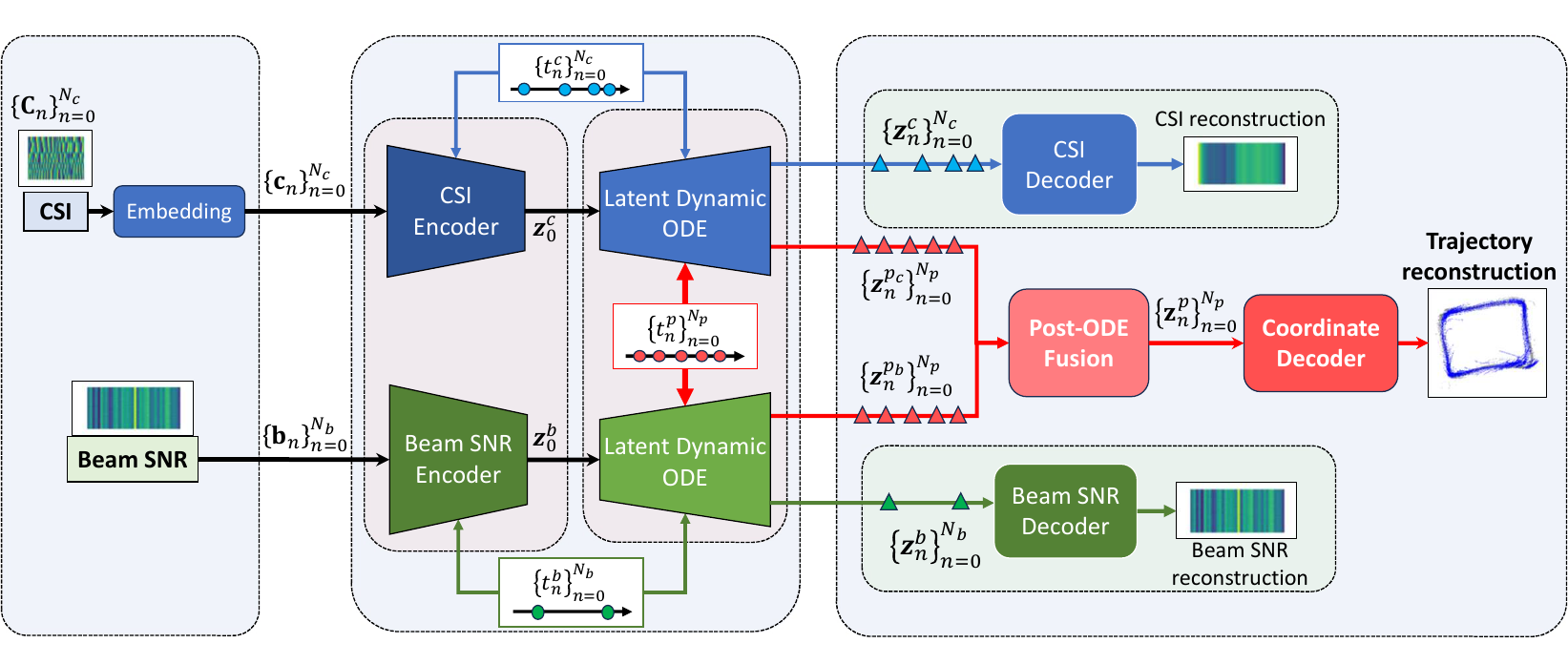}
    \vspace{-0.05in}
    \caption{The network architecture of the multi-band Wi-Fi neural dynamic fusion (NDF) for continuous-time object trajectory estimation. It comprises multiple encoders, latent dynamic learning modules, a post-ODE fusion module, and multiple decoders.}
    \label{fig:overview}
\end{figure*}

We formulate the trajectory estimation as a continuous regression problem with asynchronous {CSI} and beam SNR sequences. As illustrated in Fig.~\ref{fig:overview}, at each time instance $t_n^b$, we collect a set of $M_b$ beam SNR values $\bbf_n = [b_{n,1}, b_{n,2}, \cdots, b_{n,{M_b}}]^\top \in \mathbb{R}^{M_b \times 1}$, each corresponding to one beam training pattern. At time instance $t_n^c$, we collect a CSI measurement $\Cbf_n \in \mathbb{C}^{N_{\rm Tx} \times N_{\rm Rx} \times N_{\rm s}}$ with the $(i, j, k)$-th element $C_n(i,j,k)$  given by the CFR from transmitting antenna $i$, receiving antenna $j$ and subcarrier $k$. 
For a time window size or sequence length $\Delta T_w$, we group $N_b$ beam SNRs and $N_c$ {CSI} measurements as two input sequences. 
The problem of interest is to estimate the object trajectory $\pbf_n$ at $N_p$ desired time instances $t^p_n$ within the time window $\Delta T_w$,
\begin{equation}
    \{\bbf_n, t^b_n\}^{N_b}_{n=0}, \ \  \{\Cbf_n, t^c_n\}^{N_c}_{n=0} \rightarrow \left\{\pbf_n, t^p_n\right\}^{N_p}_{n=0},
\end{equation}
where $\pbf_n = [x_n, y_n]^\top$ consists of two-dimensional coordinates at $t^p_n$.

We follow standard practices to calibrate the raw CSI measurements $\Cbf_n$ due to the lack of synchronization between the Wi-Fi transmitter and receiver \cite{kotaru2015spotfi,  guo2017wifi, wang2017phasebeat, zeng2018fullbreathe, zhang2022practical, zhang2020calibrating}. Specifically, we use SpotFi~\cite{kotaru2015spotfi} to remove linear phase offsets caused by sampling time offset (STO) and apply an antenna-wise conjugate multiplication~\cite{zeng2018fullbreathe} to minimize packet-to-packet phase fluctuation. Once the CSI measurements are calibrated, we employ a pretrained convolutional autoencoder (CAE) to compress each calibrated CSI measurement into a CSI embedding vector $\cbf_n \in \mathbb{R}^{M_c \times 1}$, where $M_c$ is the dimension of the CSI embedding. More details about the CSI calibration and embedding can be found in Appendix~\ref{sec:csiEmbedding}. 

As illustrated in Fig.~\ref{fig:overview}, the equivalent input sequences become $\cbf_n$ and $\bbf_n$, and the problem of interest reduces to
\begin{equation}
    \{\bbf_n, t^b_n\}^{N_b}_{n=0}, \ \  \{\cbf_n, t^c_n\}^{N_c}_{n=0} \rightarrow \left\{\pbf_n, t^p_n\right\}^{N_p}_{n=0}.
\end{equation}

\subsection{Existing Solutions}
\label{sec:solutions}

\subsubsection{Frame-to-Frame Fusion}
Given the time disparity between $t^c_n$ and $t^b_n$, a simple way to combine the two input sequences is to align the CSI and beam SNR sequences at the input level. This can be accomplished using either linear or nearest-neighbor interpolation. 

For the linear interpolation (\textbf{LinearInt}) scheme, for a given output time instance $t^p_n$, we first identify the intervals $i_c$ and $i_b$ in the CSI and beam SNR sequences, respectively,  such that $t^c_{i_c}\leq t^p_n \leq t^c_{i_c+1}$ and $t^b_{i_b}\leq t^p_n \leq t^b_{i_b+1}$ and then interpolate the input sequence at $t^p_n$ using the two input measurements at both ends of the identified interval
\begin{align}\label{eq:linear_int}
    \bbf^{\text{Lin}}_n & = \bbf_{i_b} + \frac{t^p_n - t^b_{i_b}}{t^b_{i_b+1}-t^b_{i_b}} (\bbf_{i_b+1} - \bbf_{i_b}), \notag \\
    \cbf^{\text{Lin}}_n  & = \cbf_{i_c} + \frac{t^p_n - t^c_{i_c}}{t^c_{i_c+1}-t^c_{i_c}} (\cbf_{i_c+1} - \cbf_{i_c}),
\end{align}
where $n=0, \cdots, N_p$.  On the other hand, the nearest-neighbor interpolation (\textbf{NearestInt}) finds the  input element from the two input sequences at the time instance that is closest to the desired output time instance $t^p_n$
\begin{align}\label{eq:nearest_int}
    \bbf^{\text{Nea}}_n & = \begin{cases} 
\bbf_{i_b}, & \text{if } t^p_n - t^b_{i_b} \leq  t^b_{i_b+1}-t^p_n   \\
\bbf_{i_b+1}, & \text{otherwise, } \notag \\
\end{cases} \notag \\
    \cbf^{\text{Nea}}_n  & = \begin{cases} 
\cbf_{i_c}, & \text{if } t^p_n - t^c_{i_c} \leq  t^c_{i_c+1}-t^p_n   \\
\cbf_{i_c+1}, & \text{otherwise.} 
\end{cases}
\end{align}
Given the aligned input sequences
\begin{equation}
    \{\bbf^{\text{Lin/Nea}}_n, t^p_n\}^{N_p}_{n=0}, \ \  \{\cbf^{\text{Lin/Nea}}_n, t^p_n\}^{N_p}_{n=0},
\end{equation}
we can fuse them in a frame-to-frame fashion and regress the fused sequence to the trajectory coordinate
\begin{equation}\label{eq:interp-fusion}
    \hat{\pbf}_n = \mathcal{M}(\mathcal{F}(\cbf^{\text{Lin/Nea}}_n, \bbf^{\text{Lin/Nea}}_n)),
\end{equation}
where $\mathcal{F}$ represents a fusion scheme, e.g., concatenation or other considered options in Section~\ref{sec:fusion_scheme}, and $\mathcal{M}$ denotes a multi-layer perceptron (MLP) network. One may also try other interpolation schemes such as the spline and piecewise polynomial interpolation.

\subsubsection{Sequence-to-Sequence Fusion}
As opposed to the frame-to-frame fusion, one can use a recurrent neural network (RNN) to capture recurrently updated hidden features from the entire input sequence~\cite{HoangYuen19, sun2020wifi, YuSaad20}.
By fusing the hidden states corresponding to the CSI and beam SNR sequences, one can achieve what we refer to as the sequence-to-sequence fusion. 

For an input sequence $\{\sbf_n \}_{n=0}^{N}$ ($\sbf_n$ can be either the CSI $\cbf_n$ or beam SNR $\bbf_n$ sequence),  a standard RNN unit updates its hidden state $\hbf_{n-1}$ at time $t_{n-1}$ to $\hbf_{n}$ at time $t_{n}$
with the input measurement $\sbf_n$ at time instance $t_n$ as
\begin{equation} \label{rnn}
    \hbf_n = \mathcal{R}(\tilde{\hbf}_{n}, \sbf_n; \thetabf), \quad \tilde{\hbf}_{n} = {\hbf}_{n-1},
\end{equation}
where $\mathcal{R}$ is an Long Short-Term Memory~(LSTM)~\cite{hochreiter1997long} or Gated Recurrent Unit~(GRU)~\cite{chao2014learning} unit, and $\tilde{\hbf}_{n}$ is an auxiliary vector. 
In the standard {RNN}, it assumes that the sampling intervals $\Delta t_n = t_n - t_{n-1}$ are uniform, i.e.,  $\Delta t_1  = \cdots = \Delta t_N$. Consequently, the auxiliary vector is simply given by the previous hidden state $\tilde{\hbf}_{n} = {\hbf}_{n-1}$. Refer to Appendix~\ref{RNN-LSTM} for details on the standard LSTM unit update. 

To address irregularly sampled sequences where $\Delta t_n \neq \Delta t_{n+1}$, we consider the following sequence-to-sequence baseline methods. 
One such method is \textbf{RNN-Decay}~\cite{rubanova2019latent}, which decays the previous hidden state exponentially with respect to the time interval before being fed into the RNN unit, 
\begin{align}\label{eq:rnn_decay}
       \hbf_n = \mathcal{R}(\tilde{\hbf}_n, \sbf_n; \thetabf), \quad \tilde{\hbf}_n= \hbf_{n-1}e^{-\Delta t_{n}},
\end{align}
where the auxiliary vector $\tilde{\hbf}_n$ accounts for the irregular sampling intervals. Another method is \textbf{RNN-$\Deltabf$}~\cite{rubanova2019latent}, which accounts for the irregular sampling interval by augmenting the input 
\begin{align}\label{eq:rnn_delta}
        \hbf_n = \mathcal{R}(\tilde{\hbf}_{n}, \tilde{\sbf}_n; \thetabf), \quad \tilde{\hbf}_{n} = {\hbf}_{n-1}, \quad \tilde{\sbf}_n = [\sbf_n^\top, \Delta t_n]^\top,
\end{align}
while keeping the auxiliary vector as the previous hidden state. 

For a desired output time instance $t^p_n$, 
we first identify its immediate preceding time instances $i_c$ and $i_b$ in the CSI and beam SNR sequences as $t^c_{i_c}\leq t^p_n \leq t^c_{i_c+1}$ and $t^b_{i_b}\leq t^p_n \leq t^b_{i_b+1}$ with  $\hbf^c_{i_c}$ and $\hbf^b_{i_b}$ previously updated using either \eqref{eq:rnn_decay} or \eqref{eq:rnn_delta}. Then we propoagate $\hbf^c_{i_c}$ 
 and $\hbf^b_{i_b}$ to the output time instance $t^p_n$ as
\begin{align} \label{rnn_decay_fusion}
    \hbf^{p_c}_n \defeq \hbf^{c}_{t^p_n} & = \mathcal{R}(\hbf^c_{i_c}e^{-(t^p_n-t^c_{i_c})}, \mathbf{0}; \thetabf), \notag \\
    \hbf^{p_b}_n \defeq \hbf^{b}_{t^p_n} & = \mathcal{R}(\hbf^b_{i_b}e^{-(t^p_n-t^b_{i_b})}, \mathbf{0}; \thetabf), 
\end{align}
for the RNN-Decay update and 
\begin{align} \label{rnn_delta_fusion}
    \hbf^{p_c}_n \defeq \hbf^{c}_{t^p_n} & = \mathcal{R}(\hbf^c_{i_c}, [\mathbf{0}^\top, (t^p_n-t^c_{i_c})]^\top ; \thetabf),  \notag \\
  \hbf^{p_b}_n \defeq \hbf^{b}_{t^p_n}  & = \mathcal{R}(\hbf^b_{i_b}, [\mathbf{0}^\top, (t^p_n-t^b_{i_b})]^\top ; \thetabf), 
\end{align}
for the RNN-$\Delta$ update, where $n=1, \cdots, N_p$, by setting the current input $\sbf_n=\mathbf{0}$ at $t^p_n$. We can then fuse the aligned hidden states $\hbf^{p_c}_n$ and $\hbf^{p_b}_n$ at $t^p_n$ and regress the trajectory coordinate
\begin{equation} \label{eq:rnn-fusion}
    \hat{\pbf}_n = \mathcal{M}(\mathcal{F}(\hbf^{p_c}_n, \hbf^{p_b}_n)),
\end{equation}
where $\mathcal{F}$ represents a fusion scheme and $\mathcal{M}$ is an MLP.

\section{Multi-Band Neural Dynamic Fusion}
\label{sec:ndf}

In the following, we provide a detailed module-by-module explanation of the multi-band NDF framework illustrated in Fig.~\ref{fig:overview}. This framework utilizes separate encoders to sequentially map the CSI embedding $\cbf_n$ and beam SNR $\bbf_n$ sequences (along with their respective time stamps $t^c_n$ and $t^b_n$) into the latent space and estimate initial latent conditions, i.e., $\zbf^c_0$ and $\zbf^b_0$. 
In the latent dynamic learning module, both initial latent conditions are propagated using a learnable ODE model \cite{chen2018neural, rubanova2019latent}, generating virtual latent states (i.e., $\zbf^{p_c}_n$ and $\zbf^{p_b}_n$) at the same time instances $t^p_n$ for alignment.  These aligned latent states are then fused in a {post-ODE} fashion before being fed into a coordinate decoder for trajectory estimation. 
Meanwhile, one can also utilize the same learnable ODE model to regress latent states (i.e., $\zbf^{c}_n$ and $\zbf^{b}_n$) at the input time instances $t^c_n$ and $t^b_n$ for the CSI and beam SNR. These regressed latent states can be fed into either the CSI or beam SNR decoder for waveform reconstruction. 

\subsection{Encoders: An Estimator for Initital Latent Conditions} \label{sec:latent_dynamics_learning}

The purpose of the encoder is to obtain the posterior distribution of an initial latent condition corresponding to an input sequence. 
We will first consider the CSI encoder.

We start by reversing the input sequence $\cbf_{N_c},  \cdots, \cbf_0$
from the last time instance $t^c_{N_c}$ towards the initial time instance $t_0$. Then,  we map $\cbf_n$ into a hidden vector $\hbf_n^c \in \mathbb{R}^{H_c \times 1}$ with the help of an auxiliary vector $\tilde{\hbf}^c_n$
\begin{equation}\label{rnnUpdate}
    \begin{split}
        \hbf^c_n &= \mathcal{R}(\tilde{\hbf}^c_n, \cbf_n; \thetabf_g^c),
    \end{split}
\end{equation}
where $\mathcal{R}$ can be either {GRU} or {LSTM} unit with learnable parameters $\thetabf_{g}^c$.

\begin{figure}[t]
    \centering
    \includegraphics[width=1\hsize]{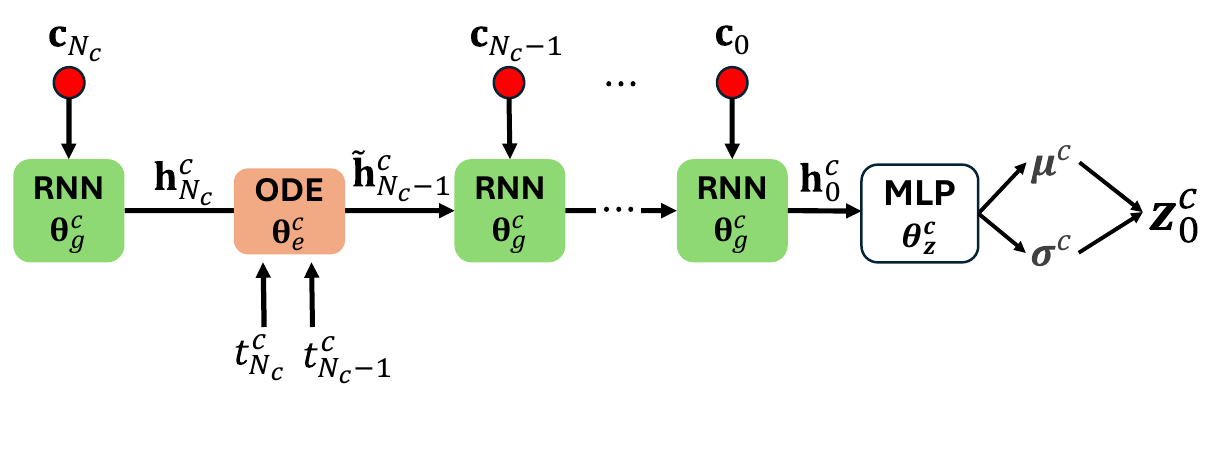}
    \caption{CSI encoder module for initial condition corresponding to the latent trajectory of the input sequence. The beam SNR encoder is similarly defined.}
    \label{fig:encoder}
\end{figure}

To handle the temporal irregularity $\Delta t^c_n = t^c_n - t^c_{n-1} \neq \Delta t^c_{n+1}$ of the input sequence, one can utilize a numerical {ODE} solver, e.g., the Euler or Runge-Kutta solvers, to propagate the hidden vector $\hbf^c_{n+1}$ at time $t^c_{n+1}$ to the auxiliary vector $\tilde{\hbf}^c_n$ at time $t^c_n$ in Fig.~\ref{fig:encoder} \cite{KellyBettencourt20,FinlayJacobsen20, ZhuJin22, NguyenNguyen22}:
\begin{align} \label{ode-encode}
    \tilde{\hbf}^c_n & = \mathcal{S}(\mathcal{O}_e,\hbf^c_{n+1}, (t_{n+1}^c, t_{n}^c); \thetabf_e^c) \notag \\
    & =\hbf^c_{n+1} + \int_{\tau = t_{n+1}^c} ^ {t_{n}^c} \mathcal{O}_e(\hbf(\tau), \tau; \thetabf_e^c) d\tau,
\end{align}
where $\mathcal{O}_e$ is a learnable ODE function represented by a neural network with parameters $\thetabf_e^c$. 

By iterating between \eqref{rnnUpdate} and \eqref{ode-encode}, we can propagate the hidden vector from $t^c_{N_c}$ to $t_0$ and output $\hbf_0^c$, which is used to estimate the initial condition $\zbf_0^c$ in the latent space and approximate its distribution by a Gaussian distribution with mean ${\mubf}^c$ and variance $(\pmb{\sigma}^c)^2$
\begin{align} \label{approxPostCSI}
q_{\theta_c}(\zbf^c_0 | \cbf_{N_c},\cdots, \cbf_0)=q_{\theta_c}(\zbf^c_0 | \hbf^c_0) =  \mathcal{N}( {\mubf}^c, (\pmb{\sigma}^c)^2),
\end{align}
Following the variational autoencoder framework~\cite{kingma2013auto}, we infer ${\mubf}^c$ and $\pmb{\sigma}^c$ from $\hbf_0$ as
\begin{equation}
    {\mubf}^c, \pmb{\sigma}^c = \mathcal{M}(\hbf^c_0; \thetabf_z^c),
\end{equation}
with $\mathcal{M}$ denoting an MLP network with parameters $\thetabf_z^c$.
Since the initial latent condition $\zbf^c_0$ is stochastic, we sample it as 
\begin{align}\label{reparametarize}
    \zbf^c_0 & = {\mubf}^c + \pmb{\sigma}^c \odot {\epsilonbf}^c, \, {\epsilonbf}^c \sim \mathcal{N}(\mathbf{0}, \Ibf_{L_c}),
\end{align}
where $\epsilonbf^c$ is a standard Gaussian sample of dimension $L_c$ and $\odot$ represents the Hadamard product.

Similarly, we can repeat the process from Eq.(\ref{rnnUpdate}) to (\ref{reparametarize}) to approximate the posterior distribution and obtain initial latent condition corresponding to the beam SNR sequence 
\begin{align}
q_{\theta_b}&(\zbf^b_0 | \bbf_{N_b},\cdots, \bbf_0)  =  \mathcal{N}( {\mubf}^b, (\pmb{\sigma}^b)^2), \label{approxPostBSNR} \\
    \zbf^b_0 & = \pmb{\mu}^b + \pmb{\sigma}^b \odot {\epsilonbf}^b, \quad {\epsilonbf}^b \sim \mathcal{N}(\mathbf{0}, \Ibf_{L_b}),  \label{postBSNR}
\end{align}
where $L_b$ is the dimension of $\zbf^b_0$. 

\subsection{Latent Dynamic Learning Modules: Alignment in Latent Space}
\begin{figure*}[t]
    \centering
    \begin{minipage}[b]{0.47\hsize}
        \centering
        \includegraphics[width=1\hsize]{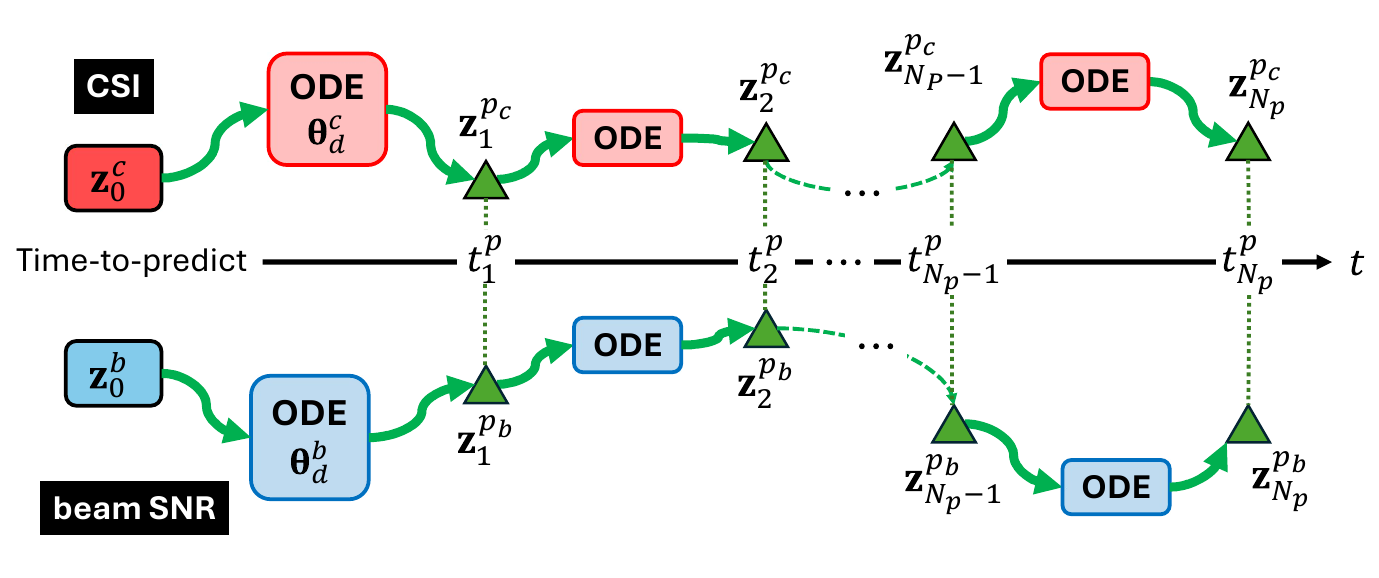}
        \subcaption{Unrolling latent dynamic ODEs for latent state alignment between CSI and beam SNR at $t^p_n$. }
    \end{minipage}
    \begin{minipage}[b]{0.5\hsize}
        \centering
        \includegraphics[width=1\hsize]{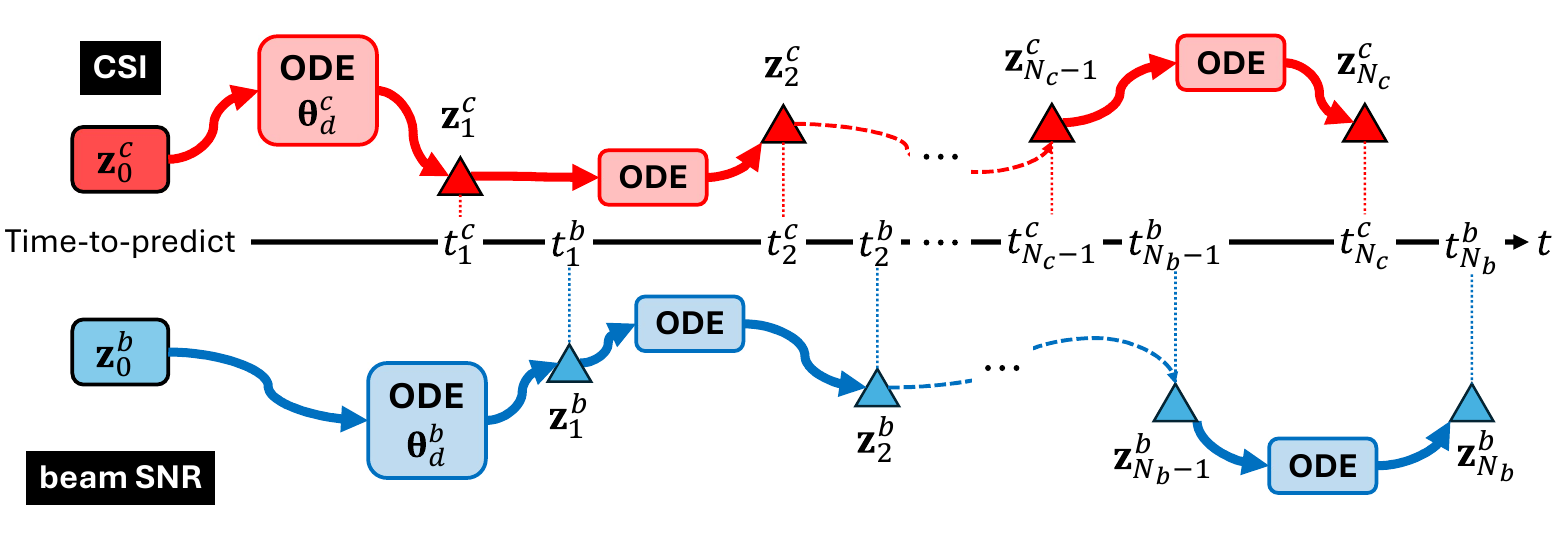}
        \subcaption{Unrolling latent dynamic ODEs for latent state recovery CSI at $t^c_n$ (top) and beam SNR at $t^b_n$ (bottom).}
    \end{minipage}
    \caption{Unrolling latent dynamic ODEs for latent state alignment (a) and latet state recovery (b).}
    \label{fig:latent_dynamic_learning}
\end{figure*}

Now, given the initial latent condition $\zbf_0$ for the input sequence, we employ a unified  continuous-time {ODE} function $\mathcal{O}_d$, modeled by a neural network with parameters $\theta_d$, to unroll the latent dynamics at any query time instance $t^q_n$. Depending on the query time instance, we have the following three cases.
\subsubsection{\texorpdfstring{$t^q_n = t^p_n$ }{} for latent state alignment} We first look into the case that the query time $t^q_n$ is the same as the time instance for coordinate estimation $t^p_n$. This is also the case that we can use the same time instance to align the latent states between the CSI and beam SNR measurements.  Specifically, we directly populate the initial latent condition to the latent state at the same query time instance for both CSI and beam SNR \cite{KellyBettencourt20,FinlayJacobsen20, ZhuJin22, NguyenNguyen22}
\begin{align}
    \zbf^{p_c}_{n} \defeq \zbf_{t_n^p}^c &= \zbf_0^c + \int_{t_0}^{t_n^p} \mathcal{O}_d (\zbf_t^c, t; \thetabf_d^c) dt, \label{align_csi} \\
    \zbf^{p_b}_{n} \defeq \zbf_{t_n^p}^b &= \zbf_0^b + \int_{t_0}^{t_n^p} \mathcal{O}_d (\zbf_t^b, t; \thetabf_d^b) dt, \label{align_bsnr}
\end{align}
where $\zbf_0^c$ and $\zbf_0^b$ are, respectively, the initial latent conditions for the CSI and beam SNR. 

In practice, we incrementally align the latent states at one query time instance at a time and then use the aligned latent states to calculate the next latent states at the next query time $t^q_{n+1}$. This is illustrated in Fig.~\ref{fig:latent_dynamic_learning} (a), where we resort to align the latent states at the first time instance $t^p_1$ for the respective initial conditions $\zbf^c_0$ and  $\zbf^b_0$,
\begin{align}
    \zbf^{p_c}_{1} &= \zbf_0^c + \int_{t_0}^{t_1^p} \mathcal{O}_d (\zbf_t^c, t; \thetabf_d^c) dt,\notag \\
     \zbf^{p_b}_{1} &= \zbf_0^b + \int_{t_0}^{t_1^p} \mathcal{O}_d (\zbf_t^b, t; \thetabf_d^b) dt, \notag
\end{align}
and then 
\begin{align}
    \zbf^{p_c}_{2} &= \zbf_1^{p_c} + \int_{t_1^p}^{t_2^p} \mathcal{O}_d (\zbf_t^c, t; \thetabf_d^c) dt, \notag \\
    \zbf^{p_b}_{2} &= \zbf_1^{p_b} + \int_{t^p_1}^{t_2^p} \mathcal{O}_d (\zbf_t^b, t; \thetabf_d^b) dt, \notag 
\end{align}
where the neural networks to represent latent ODE functions for $\zbf^c_t$ and $\zbf^b_t$ are parameterized by $\thetabf^c_d$ and $\thetabf^b_d$, respectively. 

\subsubsection{\texorpdfstring{$t^q_n = t^c_n$}{} for latent state recovery of CSI}
Similarly to the above case, we can recover the latent states of the CSI measurements at their original time instances $t^c_n$
\begin{align}
    \zbf^{c}_{n} \defeq \zbf_{t_n^c}^c &= \zbf_0^c + \int_{t_0}^{t_n^c} \mathcal{O}_d (\zbf_t^c, t; \thetabf_d^c) dt. \label{zc}
\end{align}

\subsubsection{\texorpdfstring{$t^q_n = t^b_n$}{} for latent state recovery of beam SNR}
The last case is to recover  the latent states of the beam SNR measurements at their original time instances $t^b_n$
\begin{align}
    \zbf^{b}_{n} \defeq \zbf_{t_n^b}^b &= \zbf_0^b + \int_{t_0}^{t_n^b} \mathcal{O}_d (\zbf_t^b, t; \thetabf_d^b) dt. \label{zb}
\end{align}
Note that we set that $t_0 \leq \min(t_0^c, t_0^b, t_0^p)$. In other words, the time instance for the initial latent conditions are prior to the time instance of the first measurement, either CSI or beam SNR, even before the start of the time window $\Delta T$.
More details of setting $t_0$ can be found in Sec.~\ref{sec:evaluation}.

\subsection{Post-ODE Latent Fusion}\label{sec:fusion_scheme}

Once the latent states between CSI and beam SNR measurements are aligned at $\{t^p_n\}_{n=1}^{N_p}$ in \eqref{align_csi} and \eqref{align_bsnr}, one can fuse them together as $\zbf_n^p \in \mathbb{R}^{L_p}$ of dimension $L_p$ to support the subsequent coordinate estimate. 
In the following, we consider three multi-band fusion schemes. 

\subsubsection{MLP Fusion}
\begin{figure*}[t]
    \centering
    \begin{minipage}[b]{0.32\hsize}
        \includegraphics[width=1\hsize]{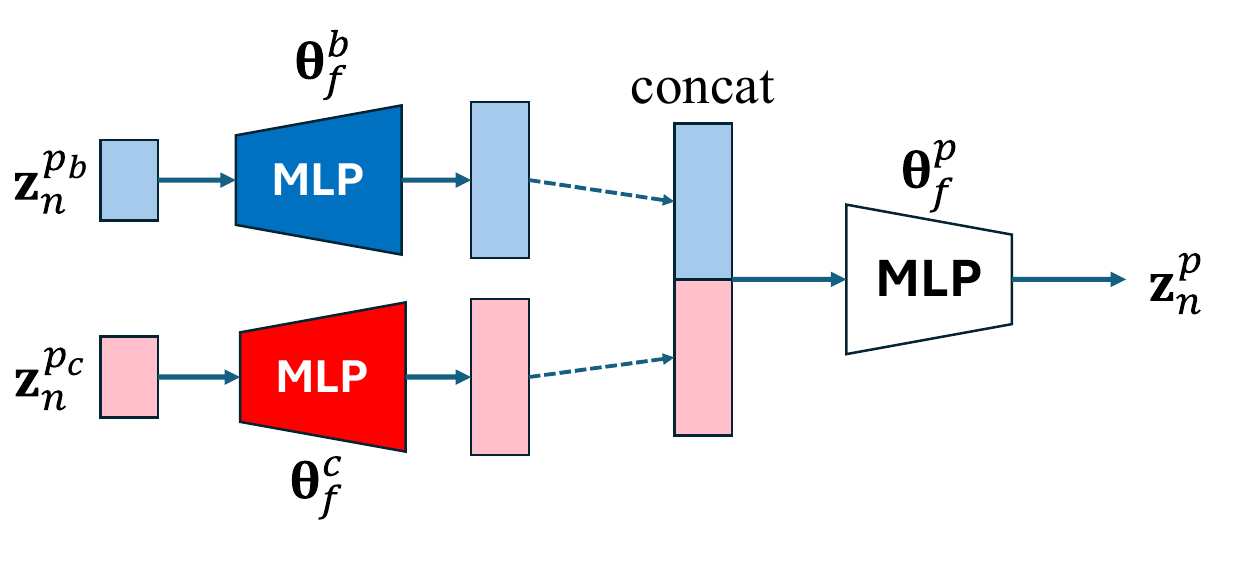}
        \subcaption{MLP Fusion}
        \label{fig:fusion_scheme_mlp1}    
    \end{minipage}
    \begin{minipage}[b]{0.28\hsize}
        \includegraphics[width=1\hsize]{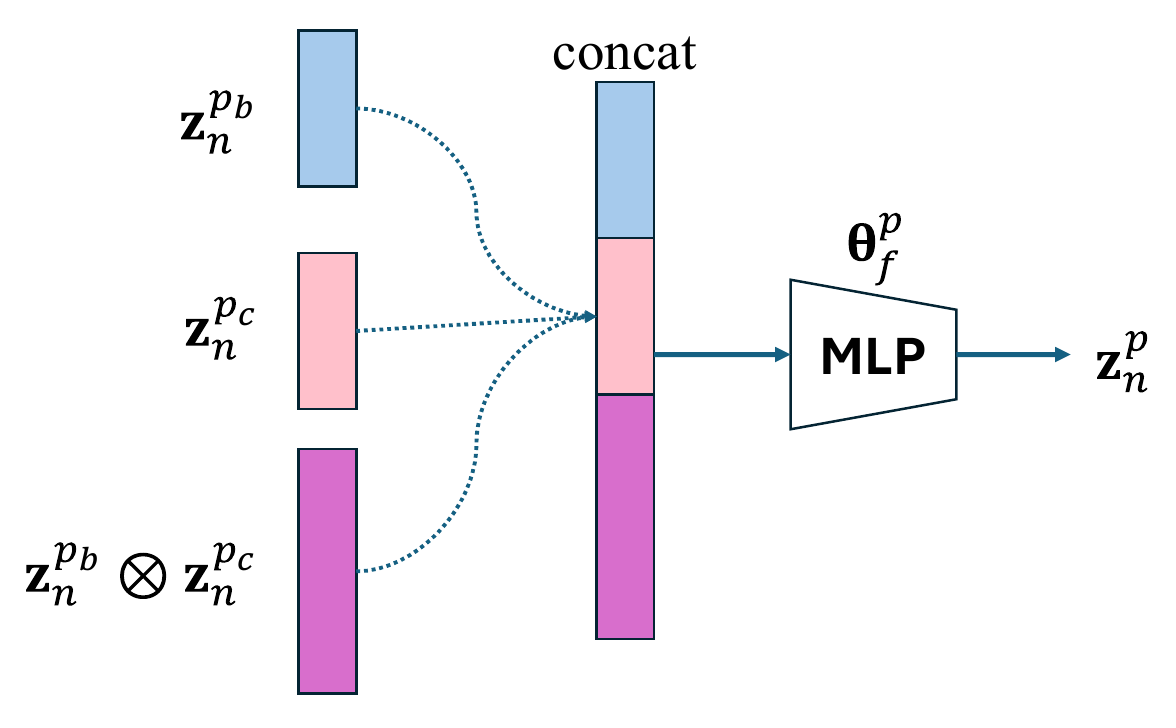}
        \subcaption{Pairwise Interaction (PI) Fusion}
        \label{fig:fusion_scheme_mlp2}    
    \end{minipage}
    \begin{minipage}[b]{0.32\hsize}
        \includegraphics[width=1\hsize]{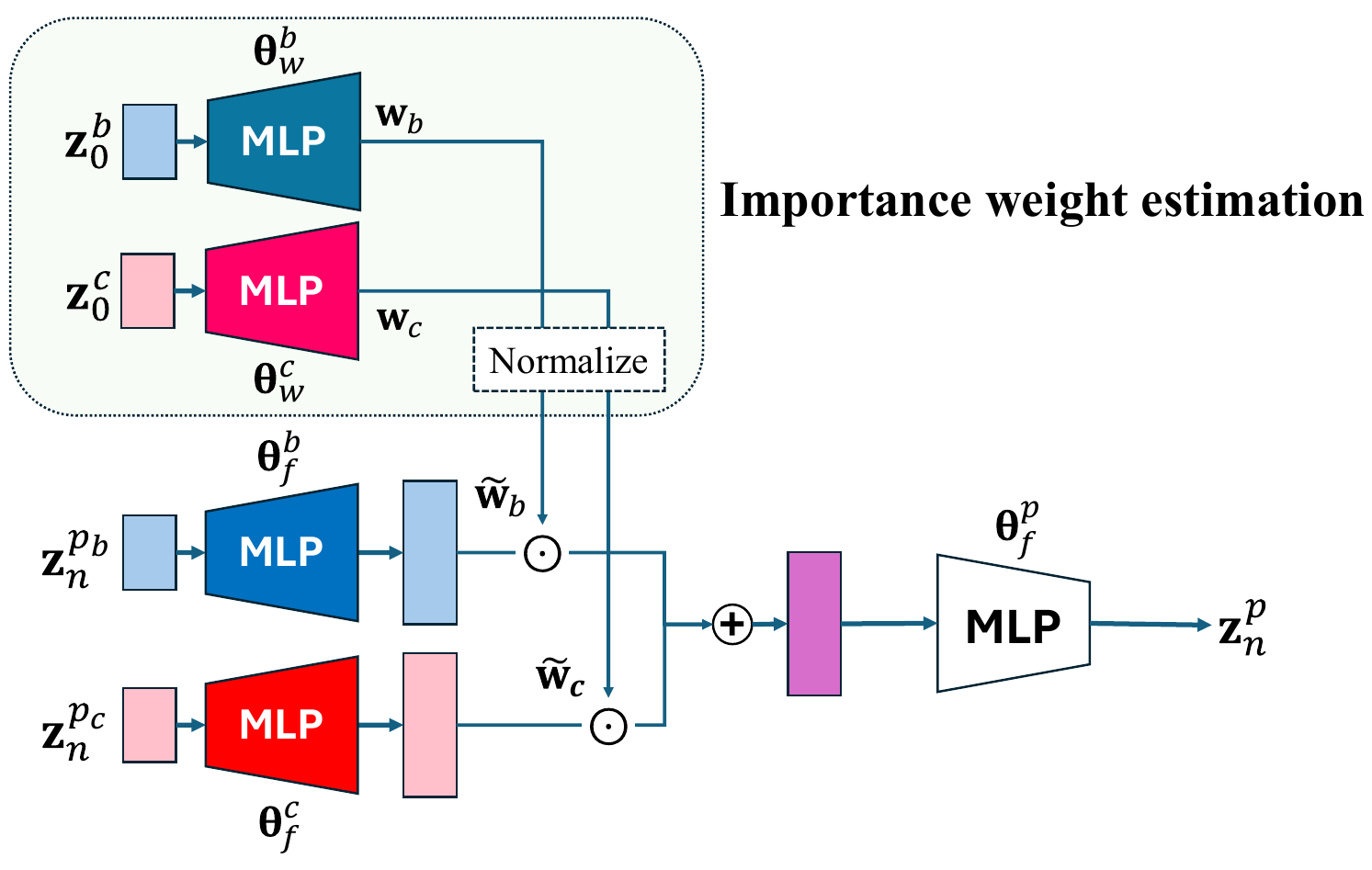}
        \subcaption{Weighted Importance (WI) Fusion}
        \label{fig:fusion_scheme_mlp3}    
    \end{minipage}
    \caption{Schemes for the post-ODE fusion.}
    \label{fig:fusion_scheme}
\end{figure*}
As shown in Fig.~\ref{fig:fusion_scheme} (a), the MLP fusion scheme starts by lifting the aligned latent states $\zbf_n^{p_c}$ and $\zbf_n^{p_b}$ to a higher dimension of $L_f$ via separate MLP networks and then projecting the concatenated latent state to a fused latet state ${\zbf}^p_n$ as
\begin{align} \label{MLPfusion}
    {\zbf}^p_n = \mathcal{M}\left( \mathcal{M}(\zbf_n^{p_b}; \thetabf_{f}^b) \oplus \mathcal{M}(\zbf_n^{p_c}; \thetabf_{f}^c); \thetabf_f^p \right), 
\end{align}
where $\thetabf_f^b$, $\thetabf_f^c$, and $\thetabf_f^p$ are learnable parameters for the three MLP networks, and $\oplus$ denotes the vector concatenation, and ${\zbf}^p_n$ is of dimension $L_p$. 

\subsubsection{Pairwise Interaction Fusion}
We combine the two aligned latent states, i.e., $\zbf^{p_c}_n$ and $\zbf^{p_b}_n$, along with their pairwise interaction $\zbf^{p_b}_n \otimes \zbf^{p_c}_n \in \Rset^{L_b L_c \times 1}$, and feed the expanded multi-band latent states to an MLP for fusion, 
\begin{align} \label{PIfusion}
    {\zbf}^p_n = \mathcal{M}\left({\zbf}_n^{p_b} \oplus {\zbf}_n^{p_c} \oplus ({\zbf}_n^{p_b} \otimes {\zbf}_n^{p_c}); \thetabf_f^p \right),  
\end{align}
where $\otimes $ denotes the Kronecker product, and $\thetabf_f^p$ represent the MLP parameters. The Kronecker term $\zbf^{p_b}_n \otimes \zbf^{p_c}_n$ accounts for cross-modal nonlinearity by expanding the dimension from $L_b$ or $L_c$ to $L_b L_c$ and including all possible element-wise multiplications between the two latent states. This is illustrated in Fig.~\ref{fig:fusion_scheme} (b). 

\subsubsection{Weighted Importance Fusion}
We also consider a weighted fusion between the two aligned latent states with their respective importance estimated directly from their initial latent conditions. This is illustrated in Fig.~\ref{fig:fusion_scheme} (c). Specifically, we first convert the initial latent states $\zbf_0^b, \zbf_0^c$ to importance weight vectors of the same dimension $L_f$:
\begin{align}
    \wbf_b = \mathcal{M}(\zbf_0^b; \thetabf_w^b)\in \mathbb{R}^{L_f}, \wbf_c = \mathcal{M}(\zbf_0^c; \thetabf_w^c)\in \mathbb{R}^{L_f},
\end{align}
where $\thetabf_w^b$ and $\thetabf_w^b$ are learnable parameters.
Then, we apply the softmax on $[\wbf_b, \wbf_c] \in \Rset^{L_f \times 2}$ over each row such that 
\begin{align}
[\tilde{\wbf}_b, \tilde{\wbf}_c] \defeq \sigma ([\wbf_b, \wbf_c]) \longrightarrow \tilde{\wbf}_b+ \tilde{\wbf}_c =\mathbf{1}_{L_f}
\end{align}
where $\sigma(\cdot)$ denotes the softmax for importance weight normalization and $\mathbf{1}_{L_f}$ is the all-one vector of dimension $L_f$. 
Meanwhile, we lift the aligned latent states $\zbf_n^{p_c}$ and $\zbf_n^{p_c}$ to a space of  dimension $L_f$ and fuse them by weighting corresponding normalized importance weights as
\begin{align} \label{WIfusion}
    \zbf_n^p = \mathcal{M}\left( [ \mathcal{M}(\zbf_n^{p_b}; \thetabf_f^b) \odot \tilde{\wbf}_b ] + [  \mathcal{M}(\zbf_n^{p_c}; \thetabf_f^c) \odot\tilde{\wbf}_c]; \thetabf_f^p \right),
\end{align}
where $\thetabf_f^b$, $\thetabf_f^c$, and $\thetabf_f^p$ are MLP parameters. 

\subsection{Decoders for Trajectory Estimation and Input Sequence Reconstruction}
\label{Decoder}
In Fig.~\ref{fig:overview}, the NDF consists of three decoders: one for estimating trajectory coordinates and the other two for reconstructing the CSI embedding and beam SNR sequences. 

\subsubsection{Trajectory Decoding}
Given the fused latent state ${\zbf}^p_n$ at desired time instances $t^p_n$, we simply employ an MLP network $\mathcal{M}$ parameterized by $\thetabf_p$ as a coordinate estimation decoder for estimating trajectory coordinates at $t^p_n$
\begin{align}\label{decode_trajectory}
    \hat{\pbf}_{n} = \mathcal{M} (\zbf_{n}^p; \thetabf_p), \quad n=0, \cdots, N_p,
\end{align}
where $\hat{\pbf}_n = [\hat{x}_n, \hat{y}_n]^T$ is the coordinate estimate at $t^p_n$ and $\thetabf_p$ is shared over all time instances of $t^p_n$.

\begin{figure}[t]
    \centering
    \includegraphics[width=0.9\hsize]{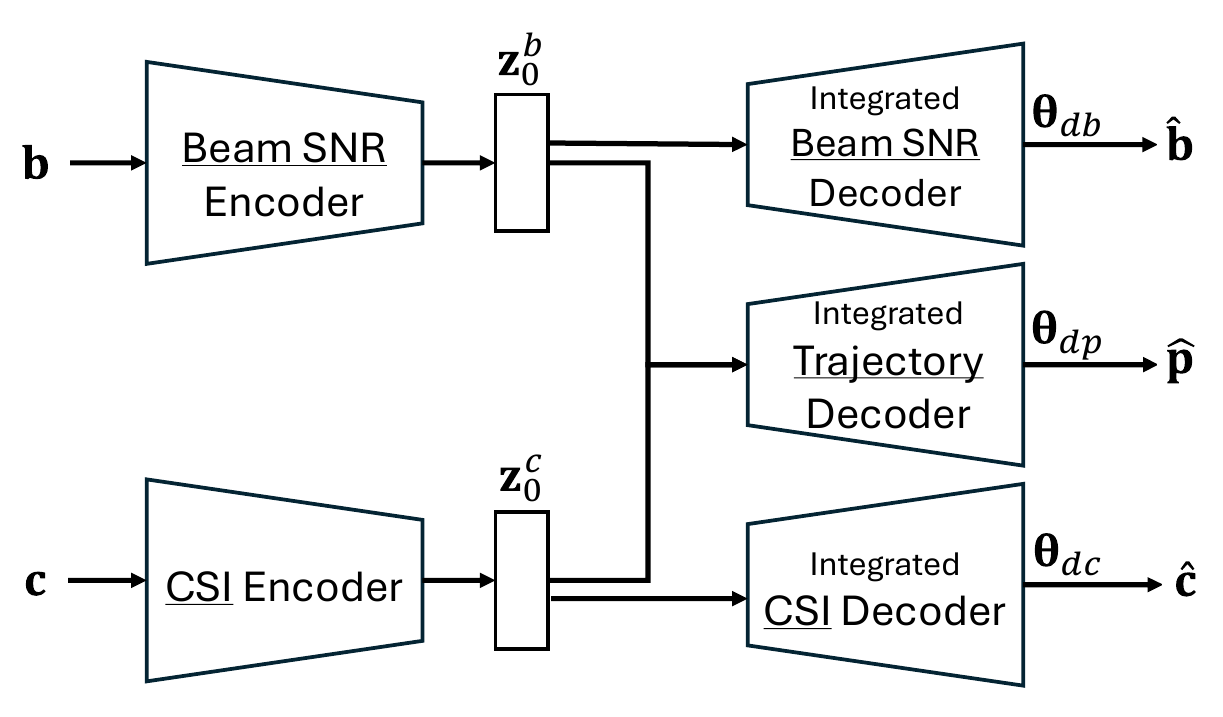}
    \caption{An equivalent multi-encoder, multi-decoder NDF structure with multiple integrated decoders directly connecting the two initial latent conditions ($\zbf_0^c$ and $\zbf_0^b$) to estimated coordinates  $\hat{\pbf}_n$ and reconstructed input sequences $\hat{\bbf}_n$ and $\hat{\cbf}_n$. }
    \label{fig:decoding_fig_loss_function}
\end{figure}

By combining the latent dynamic learning modules of \eqref{align_csi} and \eqref{align_bsnr}, the post-ODE fusion module (either \eqref{MLPfusion}, \eqref{PIfusion} or \eqref{WIfusion}), and the above trajectory decoder of \eqref{decode_trajectory}, we establish the \textbf{Integrated Trajectory Decoder} $\mathcal{P}(\cdot)$. This integrated decoder can be considered to directly take the two initial latent conditions $\zbf_0^c$ and $\zbf_0^b$ and output the coordinate estimate $\hat{\pbf}_n$
\begin{equation}
    \hat{\pbf}_n = \mathcal{P}(\zbf_0^c, \zbf_0^b, t_0, t^p_n; \thetabf_{dp}),
\end{equation}
where $\thetabf_{dp} = \{\thetabf_d^c, \thetabf_d^b, \thetabf_f, \thetabf_p\}$ with $\thetabf_f$ encompassing all learnable parameters in the post-ODE fusion model.
For instance,  $\thetabf_f =\{\thetabf^c_f, \thetabf^b_f, \thetabf^p_f \}$ for the MLP fusion, while $\thetabf_f =\{\thetabf^c_w, \thetabf^b_w,\thetabf^c_f, \thetabf^b_f, \thetabf^p_f \}$ for the weighted importance fusion.
As illustrated in Fig.~\ref{fig:decoding_fig_loss_function}, this integrated decoder structure directly links the initial latent conditions to the coordinate output and simplifies the derivation of the ELBO-based loss function in the next section. 
We hereafter group the estimated trajectory coordinates as $\hat{\pbf} = \{\hat{\pbf}_n\}_{n=0}^{N_p}$.

\subsubsection{CSI Decoding}
Given the CSI latent states $\zbf^c_n$ of \eqref{zc} at their original time instances $t^c_n$, we employ another MLP decoder with parameters $\thetabf_c$ to project $\zbf^c_n$ back to the CSI embedding sequence as
\begin{align}\label{decode_csi}
    \hat{\cbf}_{n} = \mathcal{M} (\zbf^c_n; \thetabf_c), \quad n=0, \cdots, N_c,
\end{align}
where $\thetabf_c$ is shared over all time instances of $t^c_n$. 

Similar to the integrated trajectory decoder $\mathcal{P}(\cdot)$, we combine the latent dynamic learning \eqref{align_csi} and the above CSI decoder \eqref{decode_csi} and establish the \textbf{Integrated CSI Decoder} $\mathcal{C}(\cdot)$
\begin{equation}
    \hat{\cbf}_n = \mathcal{C}(\zbf_0^c, t_0, t^c_n; \thetabf_{dc}),
\end{equation}
where $\thetabf_{dc} = \{\thetabf_d^c, \thetabf_c\}$. Equivalently, the integrated CSI decoder takes the initial latent condition $\zbf_0^c$ corresponding to the CSI input sequence and reconstructes the CSI embedding sequence as $\hat{\cbf}_n$ at $t^c_n$. 
We also group the estimated CSI embedding sequence as $\hat{\cbf} = \{\hat{\cbf}_n\}_{n=0}^{N_c}$.

\subsubsection{Beam SNR Decoding}
The last decoder is to project the latent states $\zbf^b_n$ of \eqref{zb} at $t^b_n$ back to the beam SNR sequence
\begin{align}\label{decode_bsnr}
    \hat{\bbf}_{n} = \mathcal{M} (\zbf^b_n; \thetabf_b), \quad n=0, \cdots, N_b,
\end{align}
where $\thetabf_b$ is shared over all time instances of $t^b_n$.

Combining the latent dynamic learning \eqref{align_bsnr} and the above beam SNR decoder \eqref{decode_bsnr}, we establish the \textbf{Integrated Beam SNR Decoder} $\mathcal{B}(\cdot)$ as
\begin{equation}
    \hat{\bbf}_n = \mathcal{B}(\zbf_0^b, t_0, t^b_n; \thetabf_{db}),
\end{equation}
where $\thetabf_{db} = \{\thetabf_d^b, \thetabf_b\}$.
We also group the estimated beam SNR as $\hat{\bbf} = \{\hat{\bbf}_n\}_{n=0}^{N_b}$.

\section{ELBO-Based Loss Function}

In the following, we derive an ELBO-based loss function that accounts for the multi-encoder, multi-decoder NDF architecture. 
By grouping ${\bbf}=\{{\bbf}_n\}_{n=0}^{N_b}$, ${\cbf}=\{{\cbf}_n\}_{n=0}^{N_c}$, and ${\pbf}=\{{\pbf}_n\}_{n=0}^{N_p}$ as illustrated in Fig.~\ref{fig:decoding_fig_loss_function},
The modified ELBO can be expressed as \cite{kingma2013auto}
\begin{align}\label{elbo-all}
        {\rm ELBO} =&  \mathbb{E}_{q(\zbf_0^b, \zbf_0^c|\bbf, \cbf)}[\log p(\pbf|\zbf_0^b, \zbf_0^c)] \notag \\
        & + \lambda_1 \mathbb{E}_{q(\zbf_0^b|\bbf)}[\log p(\bbf|\zbf_0^b)] + \lambda_2 \mathbb{E}_{q(\zbf_0^c|\cbf)}[\log p(\cbf|\zbf_0^c)] \notag \\
        & - \lambda_3 D_{\rm KL}[q(\zbf_0^b, \zbf_0^c|\bbf, \cbf)||p(\zbf_0^b, \zbf_0^c)] \notag \\
         \stackrel{(a)}{\approx} & \dfrac{1}{V} \sum_{v=1}^V \log p(\pbf|\zbf_0^{b(v)}\zbf_0^{c(v)}) \notag \\
        & + \dfrac{\lambda_1}{V} \sum_{v=1}^V \log p(\bbf|\zbf_0^{b(v)}) + \dfrac{\lambda_2}{V} \sum_{v=1}^V \log p(\cbf|\zbf_0^{c(v)}) \notag \\
        & - \lambda_3 D_{\rm KL}[q(\zbf_0^b, \zbf_0^c|\bbf, \cbf)||p(\zbf_0^b, \zbf_0^c)],
\end{align}
where $q(\zbf_0^c|\cbf)$ and $q(\zbf_0^b|\bbf)$ are the approximate posterior distributions defined in \eqref{approxPostCSI} and  \eqref{approxPostBSNR}, respectively, the joint posterior distribution of $\zbf_0^c$ and $\zbf_0^b$ can be factorized as 
\begin{align} \label{factorPost}
q(\zbf_0^b, \zbf_0^c|\bbf, \cbf)=q(\zbf_0^c|\cbf)q(\zbf_0^b|\bbf),
\end{align}
due to the independence assumption between the two input sequences and the use of separate encoders, $\{\lambda_i\}_{i=1}^3$ are regularization weights, $p(\zbf_0^b, \zbf_0^c)$ are the joint prior of $\zbf_0^c$ and $\zbf_0^b$ that can be also factorized as
\begin{align} \label{factorPrior}
p(\zbf_0^b, \zbf_0^c) = p(\zbf_0^b) p(\zbf_0^c),
\end{align}
with $p(\zbf_0^b)\sim {\cal{N}}(\mathbf{0}, {\Ibf}_{L_b})$ and $p(\zbf_0^c)\sim {\cal{N}}(\mathbf{0}, {\Ibf}_{L_c})$, and $p(\pbf|\zbf_0^b, \zbf_0^c)$, $p(\cbf|\zbf_0^c)$ and $p(\bbf|\zbf_0^b)$ denote the output likelihood functions of the three integrated (trajectory/CSI/beam SNR) decoders in Fig.~\ref{fig:decoding_fig_loss_function}. In the above equation, $(a)$ holds as we replace the posterior mean by its sample mean over $V$ samples of the two initial latent conditions $\zbf_0^c$ and $\zbf_0^b$ according to \eqref{reparametarize} and \eqref{postBSNR}, respectively, with $V$ independent realizations of $\epsilonbf^c$ and $\epsilonbf^b$. In practice, the number of initial latent conditions is set to $V=1$ as one can average over the independent realizations within the minibatch samples. 

For the KL divergence term $D_{\rm KL}[q(\zbf_0^b, \zbf_0^c|\bbf, \cbf)||p(\zbf_0^b, \zbf_0^c)]$, 
we invoke the independent condition between the posterior distributions of $\zbf^b_0$ and $\zbf^c_0$ given the input sequences $\bbf$ and $\cbf$ and between the prior distributions of $\zbf^b_0$ and $\zbf^c_0$
\begin{align}
        & D_{\rm KL}[q(\zbf_0^b, \zbf_0^c|\bbf, \cbf)||p(\zbf_0^b,\zbf_0^c)] \notag \\
        & \stackrel{(a)}{=} D_{\rm KL}[q(\zbf_0^b|\bbf)q(\zbf_0^c|\cbf)||p(\zbf_0^b)p(\zbf_0^c)] \notag \\
        & \stackrel{(b)}{=} D_{\rm KL}[q(\zbf_0^b|\bbf)||p(\zbf_0^b)] + D_{\rm KL}[q(\zbf_0^c|\cbf)||p(\zbf_0^c)],
\end{align}
where $(a)$ holds due to the factorization in \eqref{factorPost} and {\eqref{factorPrior}, and $(b)$ can be derived using \eqref{KLindep} in Appendix~\ref{appendix_elbo}. Then it is straightforward to show that
\begin{align}\label{elboEnd}
        D_{\rm KL} [q(\zbf_0^c|&\cbf)||p(\zbf_0^c)] = D_{\rm KL}[\mathcal{N}(\mubf^c, {\pmb{\sigma}}^c)||\mathcal{N}(\mathbf{0}, \Ibf_{L_c})] \notag \\
        &= \dfrac{1}{2} \sum^{L_c}_{l=1} \left((\mu_{l}^c)^2 + (\sigma_{l}^c)^2 - 1 - \log (\sigma_{l}^c)^2\right),  \\
        D_{\rm KL} [q(\zbf_0^b|&\bbf)||p(\zbf_0^b)] = D_{\rm KL}[\mathcal{N}(\mubf^b, {\pmb{\sigma}}^b)||\mathcal{N}(\mathbf{0}, \Ibf_{L_b})] \notag \\
        &= \dfrac{1}{2} \sum^{L_b}_{l=1} \left((\mu_{l}^b)^2 + (\sigma_{l}^b)^2 - 1 - \log (\sigma_{l}^b)^2\right),
\end{align}
where $\mu_l^{b/c}$ and $\sigma_l^{b/c}$ are the $l$-th element of $\pmb{\mu}^{b/c}$ and $\pmb{\sigma}^{b/c}$, respectively.

For output log-likelihood functions, we start with the integrated trajectory decoder $\cal{P}(\cdot)$ that takes the two initial latent conditions and estimates the trajectory coordinates at $t^p_n$,
\begin{align}\label{elbo_d_bsnr_start}
        \log p(\pbf|\zbf_0^b, \zbf_0^c) &= \log p(\pbf_0, \pbf_1, \cdots, \pbf_{N_p}|\zbf_0^b, \zbf_0^c) \notag \\
        &  \stackrel{(a)}{\approx} \sum_{n=1}^{N_p} \log p(\pbf_n|\zbf_0^b, \zbf_0^c),
\end{align}
where the approximation $(a)$ holds as we invoke an independent assumption over the sequential coordinate outputs over the time instance $n$.
We assume that each element in $\pbf_{n}=[x_n, y_n]^\top$ follows a Laplace distribution:
\begin{align}\label{laplaceDist}
    p(x_{n}|\zbf_0^b, \zbf_0^c) = & \dfrac{1}{2b_p}\exp \left(-\dfrac{|x_{n} -\hat{x}_n|}{b_p}\right), \notag \\
    p(y_{n}|\zbf_0^b, \zbf_0^c) = & \dfrac{1}{2b_p}\exp \left(-\dfrac{|y_{n} -\hat{y}_n|}{b_p}\right),
\end{align}
where $b_p \in \mathbb{R}$ is a scaling parameter and $\hat{\pbf}_{n} = [\hat{x}_n, \hat{y}_n]^\top =  \mathcal{P}(\zbf_0^c, \zbf_0^b, t_0, t^p_n; \thetabf_{dp})$ is the estimated trajectory coordinate at $t_n^p$.
As a result, we can show that
\begin{equation}
    \log p(\pbf_n|\zbf_0^b, \zbf_0^c)  \propto - \dfrac{\|\pbf_n - \hat{\pbf}_n\|_1}{b_p},
\end{equation}
where $\|\cdot\|_1$ denotes the $\ell_1$ norm.
Assuming $b_p=1$ and plugging the above equation back to \eqref{elbo_d_bsnr_start}, the output log-likelihood function of the integrated trajectory decoder is given as
\begin{equation}\label{elbo_d_bsnr_end}
    \log p(\pbf|\zbf_0^{b}, \zbf_0^{c}) \propto -\sum^{N_p}_{n=1} \|\pbf_n - \hat{\pbf}_n\|_1.
\end{equation}
It is seen that maximizing this log-likelihood is equivalent to minimizing mean absolute error (MAE) between ground truth and estimated trajectory coordinates.
We can follow Eq.~(\ref{elbo_d_bsnr_start}) to (\ref{elbo_d_bsnr_end}) for the output log-likelihood functions of the integrated beam SNR and CSI decoders,
\begin{align}
    \log p(\bbf|\zbf_0^{b}) \propto -\sum^{N_b}_{n=1} \|\bbf_n - \hat{\bbf}_n\|_1, \\
    \log p(\cbf|\zbf_0^{c}) \propto -\sum^{N_c}_{n=1} \|\cbf_n - \hat{\cbf}_n\|_1.
\end{align}

Combining the KL divergence term and the output log-likelihood functions of the integrated decoders, the modified ELBO \eqref{elbo-all} reduces to the following loss function
\begin{align}\label{eq:final_objective_function}
        \mathcal{L} =& \sum^{N_p}_{n=1} \|\pbf_n - \hat{\pbf}_n\|_1 + \lambda_1 \sum^{N_b}_{n=1} \|\bbf_n - \hat{\bbf}_n\|_1 + \lambda_2 \sum^{N_c}_{n=1} \|\cbf_n - \hat{\cbf}_n\|_1 \notag \\
        &+ \lambda_3 \sum^{L_b}_{l=1} \left((\mu_{l}^b)^2 + (\sigma_{l}^b)^2 - 1 - \log (\sigma_{l}^b)^2\right) \notag \\
        &+ \lambda_4 \sum^{L_c}_{l=1} \left((\mu_{l}^c)^2 + (\sigma_{l}^c)^2 - 1 - \log (\sigma_{l}^c)^2\right)
\end{align}
where we relax the regularization weight $\lambda_3$ for the joint KL term to different regularization weights $\lambda_3$ and $\lambda_4$ for individual KL terms of beam SNR and CSI, respectively. 

\section{Performance Evaluation}
\label{sec:evaluation}

\subsection{In-House Testbed and Data Collection}
\begin{figure}[t]
    \centering
    \includegraphics[width=1\hsize]{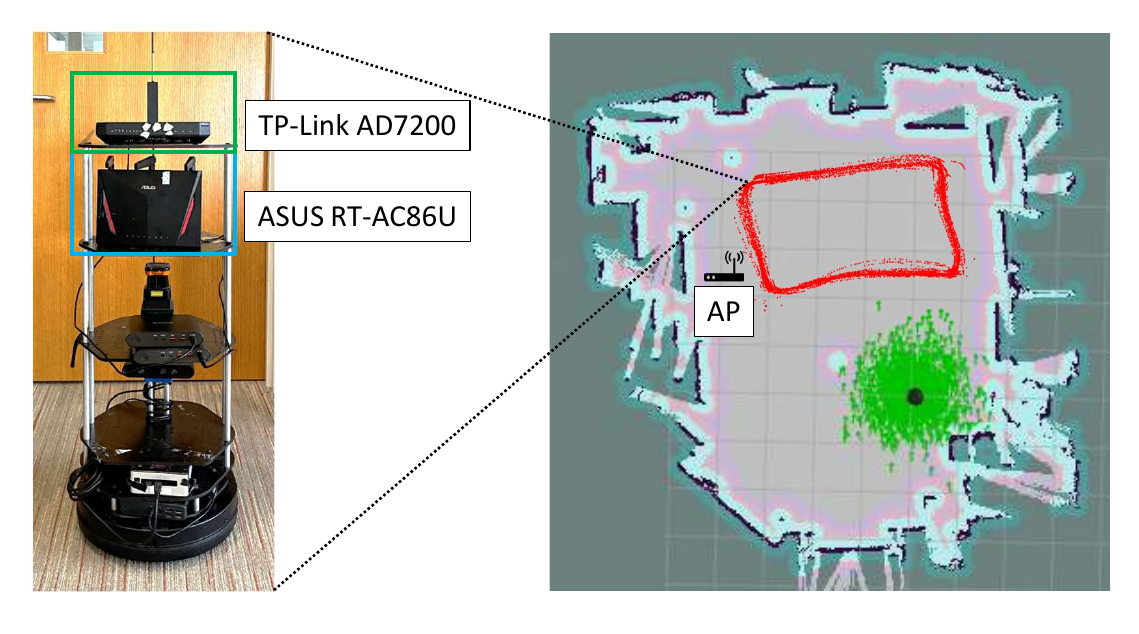}
    \caption{In-house multi-band Wi-Fi testbed: a TurtleBot with a TP-Link router for $60$-GHz beam SNR and an ASUS router for $5$-GHz CSI (left); and the data collection floorplan with the AP location and trajectories (right). }
    \label{fig:testbed_route}
\end{figure}
We upgrade our previous in-house testbed in \cite{RubioWang23} and \cite{RubioWang24} from collecting single-band beam SNRs to simultaneously gathering both $5$-GHz CSI and $60$-GHz beam SNRs.
As shown in Fig.~\ref{fig:testbed_route}, we mount two routers on a TurtleBot as a mobile user: one router is the $802.11$ac-compliant ASUS RT-AC86U device for collecting $80$-MHz CSI data at $5$ GHz and the other $802.11$ad-compliant TP-Link Talon AD7200 router for $60$-GHz beam SNR. The mobile user TurtleBot moves along predefined rectangular trajectories (denoted by red dot lines in the right plot of Fig.~\ref{fig:testbed_route}) in a large conference room. Positioned at the lower left corner of the rectangular trajectory, another pair of identical routers act as a multi-band AP. 

To enable data collection on these commercial-off-the-shelf routers, we replace the original firmware with open-source ones~\cite{GringoliSchulz19, SteinmetzerWegemer18} and follow the methods of \cite{BielsaPalacios18} and \cite{GringoliSchulz19} to extract the beam SNR and CSI from the commercial routers.
From the four antennas (three external and one internal) of the ASUS router, we are able to extract $N_{\rm Tx} \times N_{\rm Rx} = 4 \times 2$ spatial streams of CSI over $N_{\rm s}=234$ subcarriers, excluding null subcarriers. 
Each raw CSI frame $\Cbf_n \in \Cset^{4 \times 2 \times 234}$ is calibrated and  compressed into the CSI embedding input $\cbf_n \in \mathbb{R}^{36\times 1}$ with $M_c=36$, as described in Appendix~\ref{sec:csiEmbedding}. 
On the other hand, the TP-Link router employs an analog phase array of $32$ antenna elements and sequentially scans over $M_b = 36$ predefined directional beampatterns, leading to $\bbf_n \in \Rset^{36 \times 1}$.

Our testbed is also equipped with a LiDAR and a wheel encoder to self-localize over a predefined map. The self-localized coordinates, recorded at a frame rate of $10$ frames per second (fps), are then used as ground-truth labels $\pbf_n$ for trajectory estimation. The system clocks of all networked devices, including the routers and the TurtleBot, are precisely synchronized using the Network Time Protocol (NTP) with a central desktop acting as the NTP server. The desktop controls and aligns the clocks of all other devices over the network connection, ensuring that the timestamps across the network refer to the same clock. Consequently, we obtained $43,277$ frames for CSI and coordinate labels, and $9,590$ frames for beam SNRs.

\begin{figure}[t]
    \centering
    \includegraphics[width=0.9\hsize]{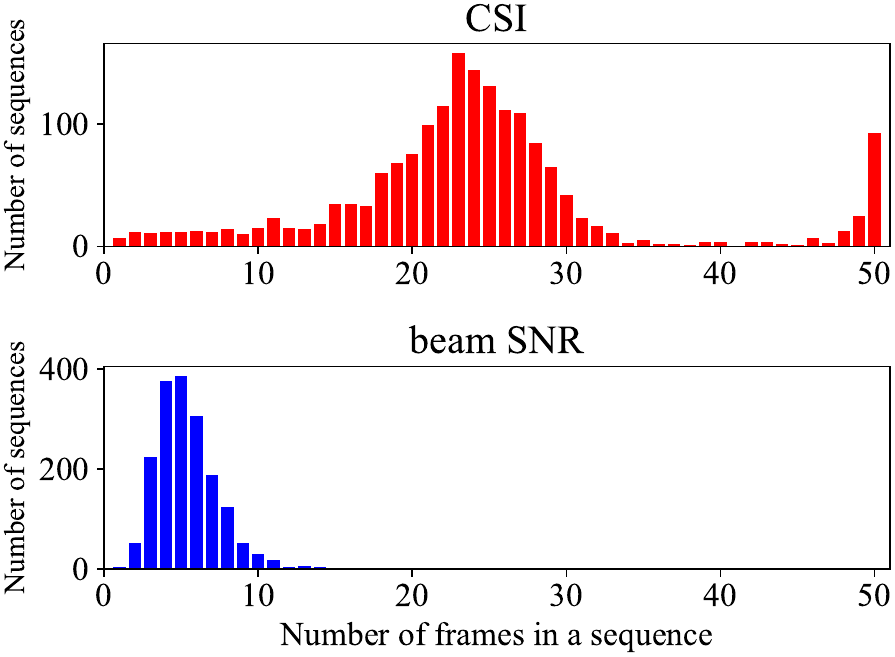}
    \caption{The histogram of the number of CSI (top) and beam SNR (bottom) frames over non-overlapping $5$-sec sequences.}
    \label{fig:samples_in_sequence}
\end{figure}

\subsection{Implementation}

We set $\Delta T_w = 5$ seconds to group all collected CSI frames, beam SNRs, and coordinate labels into sequences. Fig.~\ref{fig:samples_in_sequence} shows the histograms of the number of CSI (top) and beam SNR (bottom) frames over non-overlapping sequences of $5$ seconds. It reveals that most CSI sequences have $20-30$ frames over a period of $5$ seconds, yielding an average frame rate of $4-6$ fps. In comparision, the beam SNR sequences contain much less number of frames over $5$ seconds, yielding an avarage frame rate of $1$ fps. 
For each sequence, all timestamps $\{t_n^b\}_{n=0}^{N_b}$, $\{t_n^c\}_{n=0}^{N_c}$, and $\{t_n^p\}_{n=0}^{N_p}$ are normalized into $[0, 1]$ by dividing the relative timestamps by $5$ seconds.

We consider three data splits for performance evaluation: 
\begin{enumerate}
\item \textbf{Random Split}: We group the frames into $1,778$ non-overlapping sequences of $5$ seconds (with a stepsize of $5$ seconds) and randomly divide these non-overlapping sequences into train, validation, and test sets with a ratio of $80:10:10$. The results in the following subsections are based on this data split.
\item \textbf{Temporal Split}: 
We divide all frames into training and test sets strictly according to their chronological order. Specifically, we group the first collected $s\%$ of frames into the training set, and the remaining frames into the test set.  In other words, all test frames represent future data that was not seen during the training phase. This sequential split is used to evaluate the temporal generalization performance in Section~\ref{sec:extraplo} with different values of $s$. With a sequence length of $5$ seconds, we group the training and test frames into sequences with stepsizes of $1$ and, respectively, $5$ seconds. 
\item \textbf{Coordinate Split}: We also divide all frames into training and test sets according to their ground truth coordinates.  Specifically, we keep frames from a particular area (e.g., a corner) in the test set, completely unseen from the training set. This split is used to evaluate the generalization performance at unseen coordinates in Section~\ref{sec:unseen}, which is referred to as the spatial generalization. 
\end{enumerate}

We use an autoencoder with $3$ 1D convolutional layers and $3$ MLP layers for the pretraining discussed in Appendix~\ref{sec:csiEmbedding} to obtain the CSI embedding vector $\cbf_n$.  Both beam SNR and CSI embedding sequences are normalized to $[0, 1]$. For the encoder, we use the GRU unit with a hidden dimension of $H_b=H_c=20$ for the beam SNR and CSI input sequences. We set $L_b = L_c = 20$ for the dimensions of the initial latent condition and unrolled latent states  $\zbf^b_n$ and $\zbf^c_n$, $n=0, 1, \cdots, N_b/N_c$. We lift the aligned latent state to a space of dimension $L_f=128$ before projecting it back to the fused latent space of $L_p = 20$.  We employ the Euler and Dopri5 ODE solvers for encoding and latent dynamic ODE, respectively. The decoders for trajectory~(\ref{decode_trajectory}), CSI~(\ref{decode_csi}) and beam SNR~(\ref{decode_bsnr}) share the same MLP architecture of three MLP layers. 

The set of regularization parameters is chosen by performing a hyperparameter search in the interval of $[0, 1]$ using Optuna~\cite{takuya2019optuna}. It is based on the validation loss transition of coordinate estimation within  $125$ epochs and $100$ trials are executed.
Fig.~\ref{fig:optuna_parameter_search} illustrates the loss function as a function of regularization parameters ($\lambda_1$, $\lambda_3$) for beam SNR and ($\lambda_2$, $\lambda_4$) for CSI, where red dots denote the values of hyperparameter pairs achieving the smallest validation loss over $125$ epochs or the smallest intermediate loss if terminated in an earlier epoch. As a result, we set the regularization parameters as $\lambda_1 = 0.7, \lambda_2 = 1.0, \lambda_3 = 0.0010, \lambda_4 = 0.25$ in the ELBO-based loss function of \eqref{eq:final_objective_function}. 
To train the NDF network, we set the minbatch size to $32$ and the maximum number of epochs is $250$, and we save the model achieving the best validation loss while training.
We used the Adamax optimizer with the maximum learning rate of $4e-3$ with the OneCycle learning rate scheduling for fast convergence~\cite{SmithTopin19}. 

\begin{figure}[t]
    \centering
    \begin{minipage}[b]{0.48\hsize}
        \includegraphics[width=1\hsize]{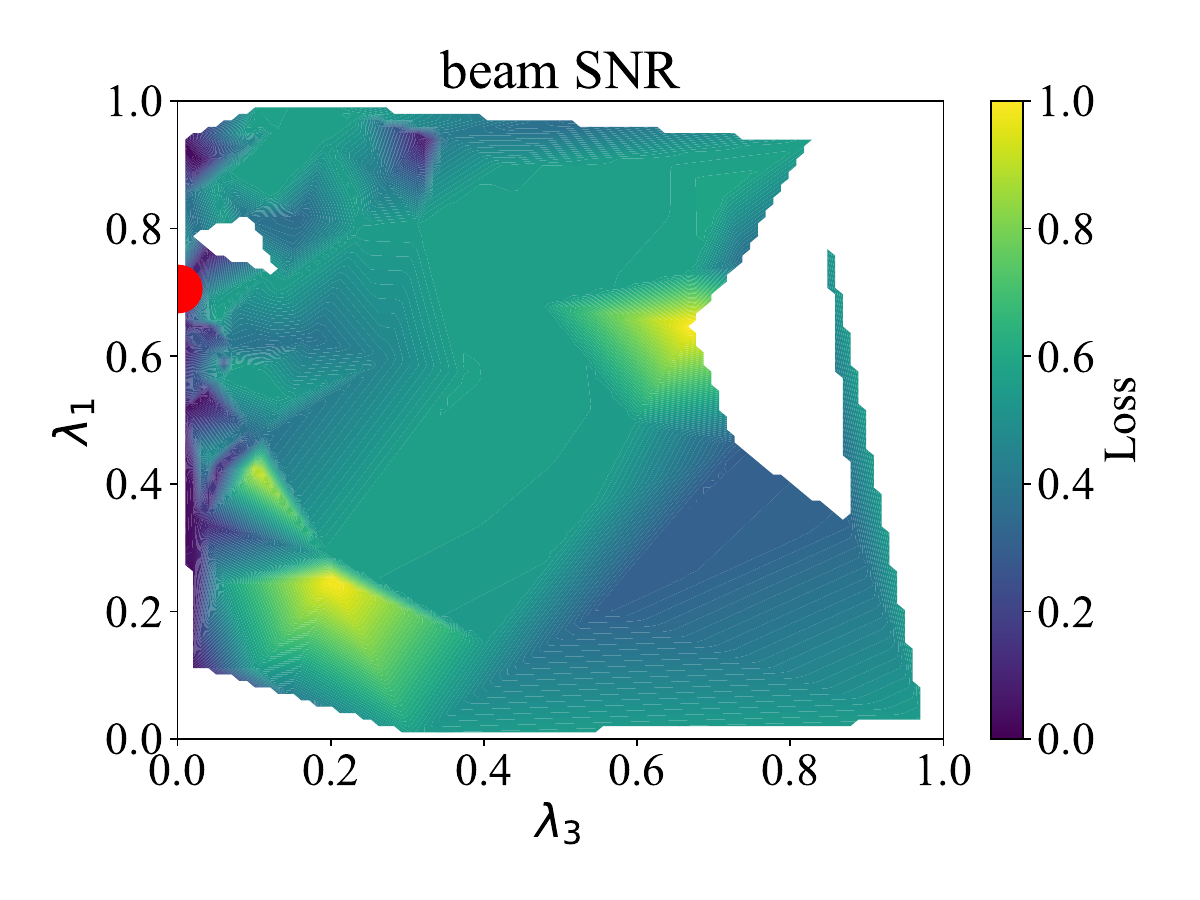}
        \subcaption{beam SNR}
    \end{minipage}
    \begin{minipage}[b]{0.48\hsize}
        \includegraphics[width=1\hsize]{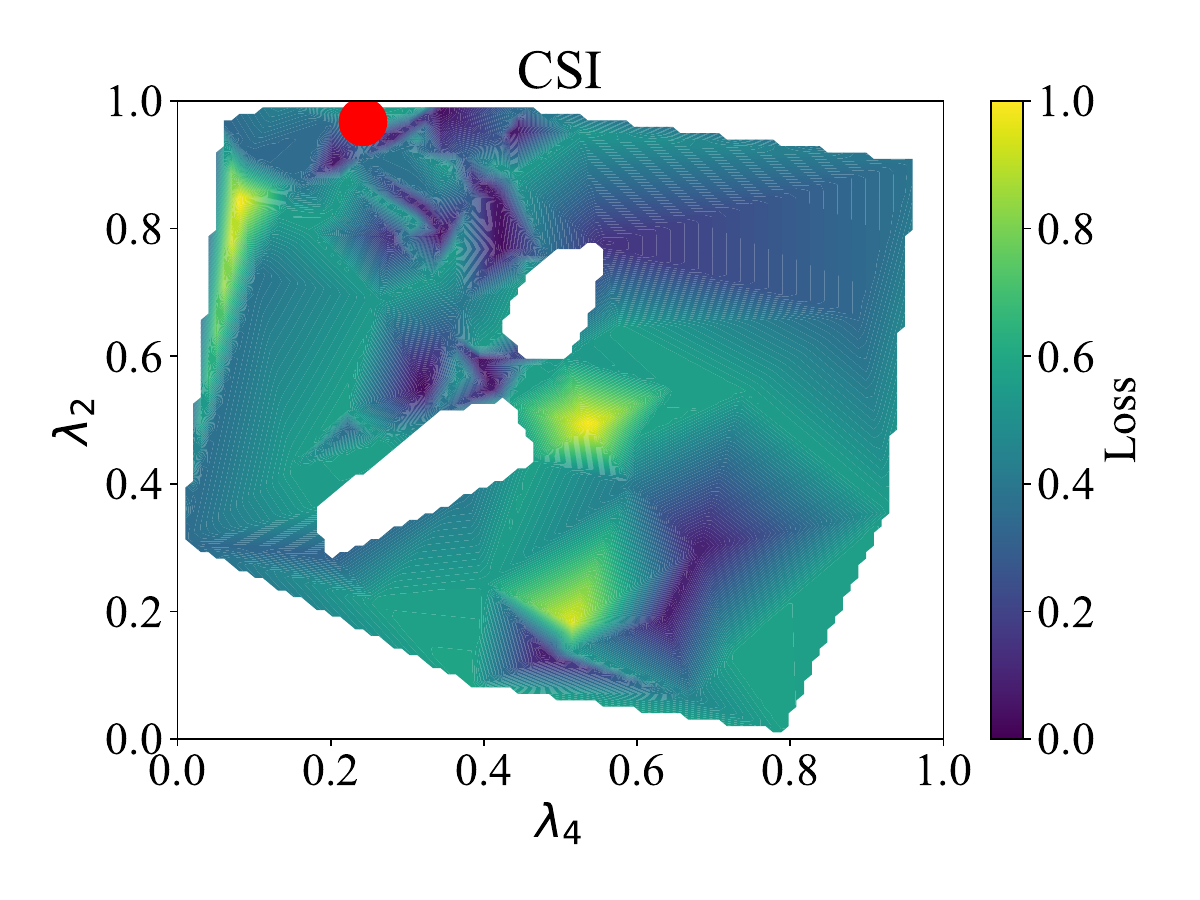}
        \subcaption{CSI}
    \end{minipage}
    \caption{The search for regularization parameters in the ELBO-based loss function of \eqref{eq:final_objective_function}.}
    \label{fig:optuna_parameter_search}
\end{figure}

\subsection{Comparison to Baseline Methods}
For performance comparison, we consider a comprehensive list of baseline methods
\begin{itemize}
    \item \textbf{Single-band} methods (either CSI or beam SNR): 1)  Linear interpolation (LinearInt) of (\ref{eq:linear_int}); 2) Nearest interpolation (NearestInt) of (\ref{eq:nearest_int}); 3) RNN-Decay of (\ref{eq:rnn_decay}); 4) RNN-$\Delta$ of \eqref{eq:rnn_delta}; 5) DDND of \cite{RubioWang23, RubioWang24}.
    \item \textbf{Frame-to-Frame Fusion} methods: 1) LinearInt fusion of \eqref{eq:linear_int} and  \eqref{eq:interp-fusion}; 2) NearestInt fusion of (\ref{eq:nearest_int}) and (\ref{eq:interp-fusion}).
    \item \textbf{Sequence-to-Sequence Fusion} methods: 1) RNN-Decay fusion of \eqref{eq:rnn_decay}, \eqref{rnn_decay_fusion} and \eqref{eq:rnn-fusion}; 2) RNN-$\Delta$ fusion of  \eqref{eq:rnn_delta}, \eqref{rnn_delta_fusion} and \eqref{eq:rnn-fusion}.
\end{itemize}

\begin{table*}[htb]
    \centering
    \caption{Localization errors (m) for all single-band and multi-band baseline and the proposed NDF methods. }
    \begin{tabular}{lccccccccc} \toprule
    & \multicolumn{6}{c}{\textbf{Single-band}} & \multicolumn{3}{c}{\textbf{Multi-band}} \\
    \cmidrule(lr){2-7} 
    & \multicolumn{3}{c}{\textbf{CSI}} & \multicolumn{3}{c}{\textbf{beam SNR}} &  \\
    \cmidrule(lr){2-4}\cmidrule(lr){5-7}\cmidrule(lr){8-10}
    & Mean & Median & CDF@0.9 & Mean & Median & CDF@0.9 & Mean & Median & CDF@0.9 \\ \midrule
    LinearInt & 1.79 & 1.89 & 2.91 & 0.932 & 0.726 & 1.77 & 0.764 & 0.506 & 1.76 \\
    NearestInt & 1.82 & 1.92 & 2.87 & 1.00 & 0.715 & 2.09 & 0.839 & 0.553 & 2.01 \\
    RNN-Decay & 1.02 & 0.739 & 2.27 & 1.03 & 0.545 & 2.64 & 0.685 & 0.432 & 1.69 \\
    RNN-$\Delta$ & 1.03 & 0.733 & 2.39 & 0.994 & 0.499 & 2.70 & 0.506 & 0.215 & 1.42 \\
    DDND & 0.975 & 0.588 & 2.45 & 0.390 & 0.191 & 0.859 & - & - & - \\
    NDF (ours) & - & - & - & - & - & - & \textbf{0.263} & \textbf{0.148} & \textbf{0.611} \\ \bottomrule
    \end{tabular}
    \label{tab:error_statistics}
\end{table*}

Table~\ref{tab:error_statistics} summarizes the trajectory estimation performance of all baseline methods and the proposed NDF method under the random sequence split. By comparing the mean, median, and the $90$th percentile of the localization error in the unit of meters, it is seen that, for a given method, e.g., the linear interpolation or the RNN-Decay, the multi-band fusion improves the localization performance from either the CSI-only or the beam SNR-only methods. Comparison between the interpolation (i.e., linear and nearest) and RNN methods (i.e., RNN-Decay and RNN-$\Delta$) shows that the RNN-based methods can significantly improve the CSI-only performance and contribute to the overall improvement using both CSI and beam SNR. If we narrow down to the last column of Table~\ref{tab:error_statistics}, it is clear that, by properly aligning the latent states using the latent dynamic ODE, the NDF can further reduce the location error from the best multi-band baseline (i.e., RNN-Decay) to a mean localization error of $14.8$ cm. Fig.~\ref{fig:performance_cdf_multiband} highlights the cumulative distribution functions (CDFs) of the localization error from the multi-band methods and the NDF. 

\begin{figure}[t]
    \centering
    \includegraphics[width=1\hsize]{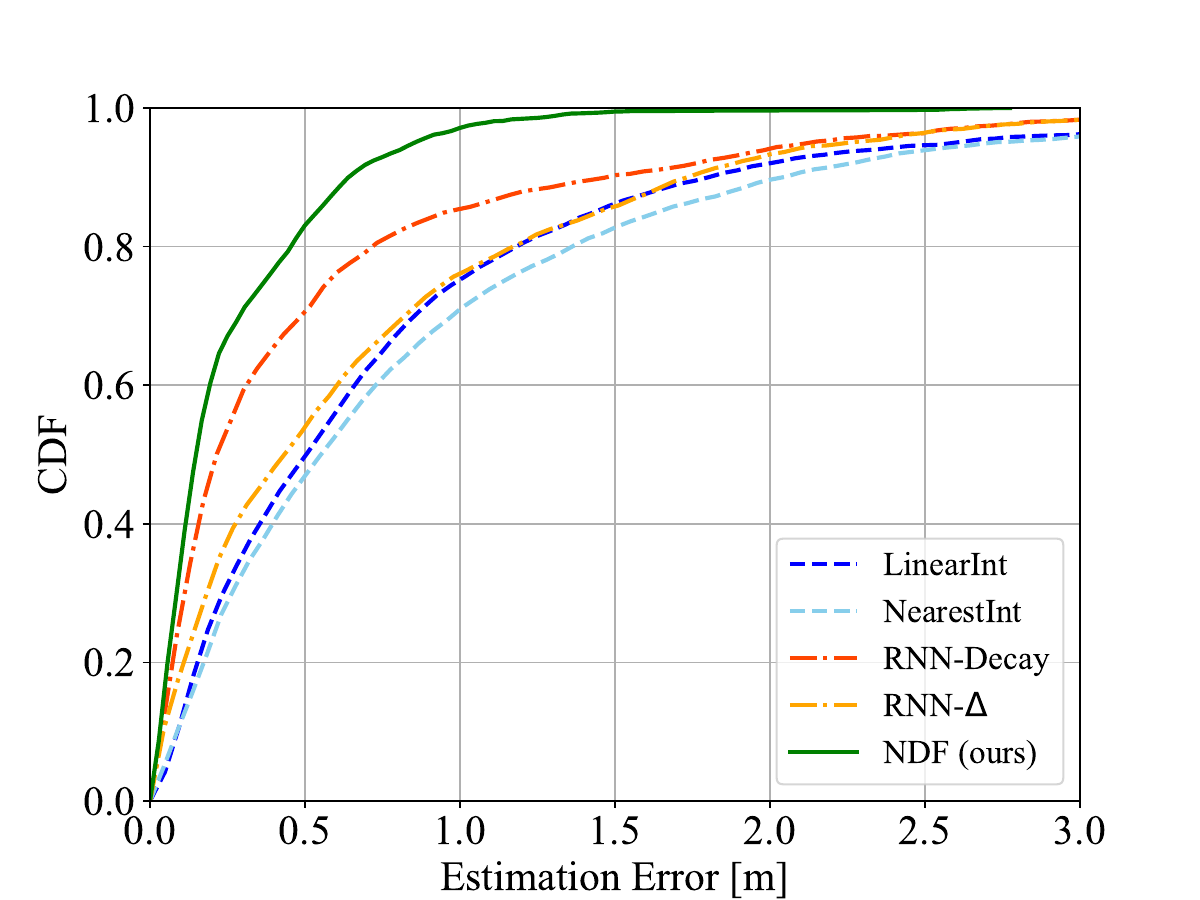}
    \caption{Cumulative distribution function (CDF) of localization errors for the multi-band fusion baseline and NDF methods.}
    \label{fig:performance_cdf_multiband}
\end{figure}

To qualitatively compare the baseline and proposed methods, we overlap the estimated trajectories (in red dots) with the groundtruth coordinates (in dim blue dots) in Fig.~\ref{fig:performance_visualization_main} for selected multi-band fusion baseline methods (nearest interpolation, RNN-Decay, RNN-$\Delta$) and the proposed NDF method. The improvement from the frame-to-frame fusion (nearest interpolation) to the sequence-to-sequence fusion (RNN-Decay, RNN-$\Delta$) is noticeable as there are less localization errors at the center of the rectangular area. The NDF shows the best results by significantly reducing the outliers and forcing the trajectory estimates along the rectangular track.

\begin{figure}[t]
    \centering
    \begin{minipage}[t]{0.475\hsize}
        \centering
        \includegraphics[width=1\hsize]{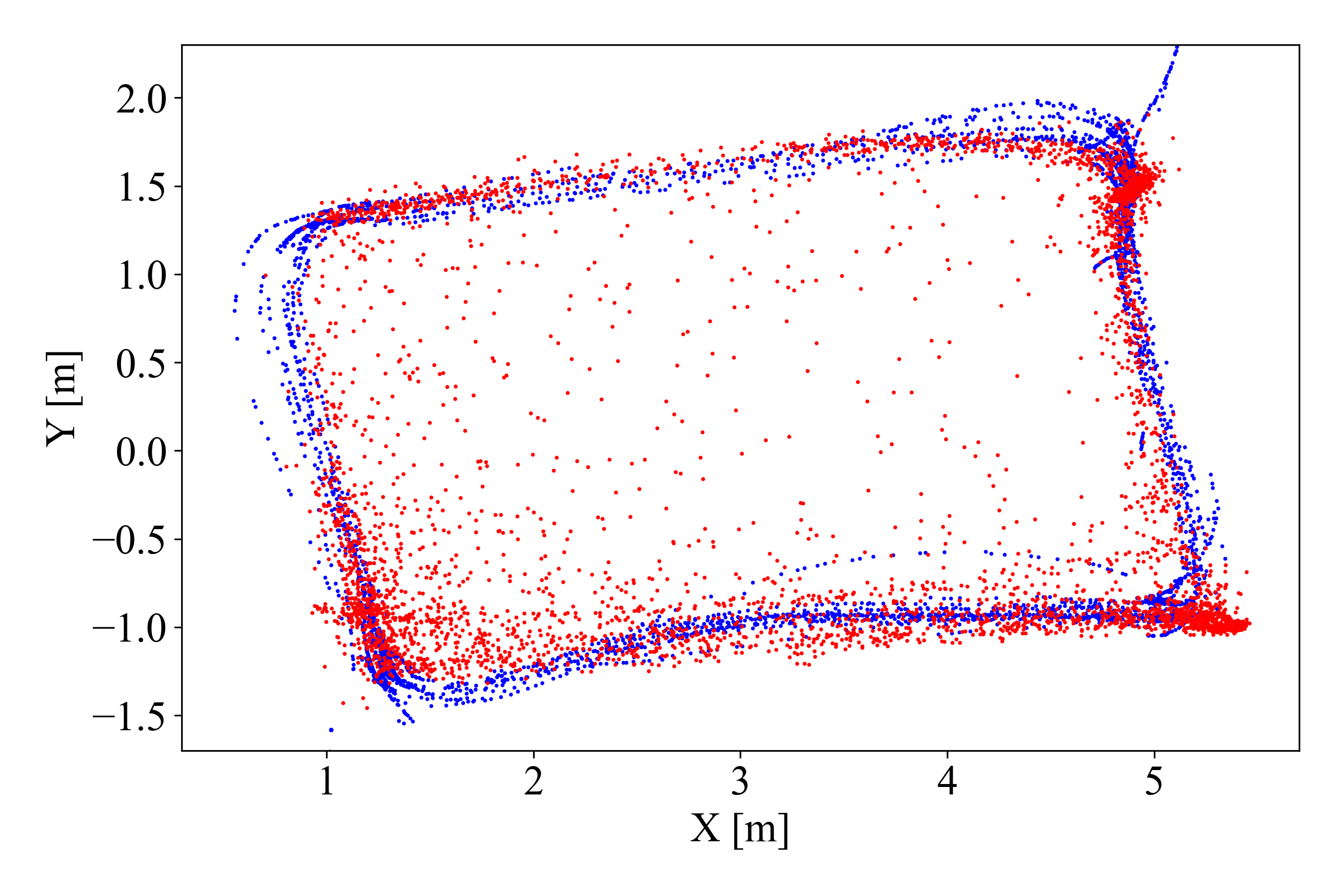}
        \subcaption{NearestInt}\label{fig:estimate_nearest}
    \end{minipage}
    \begin{minipage}[t]{0.475\hsize}
        \centering
        \includegraphics[width=1\hsize]{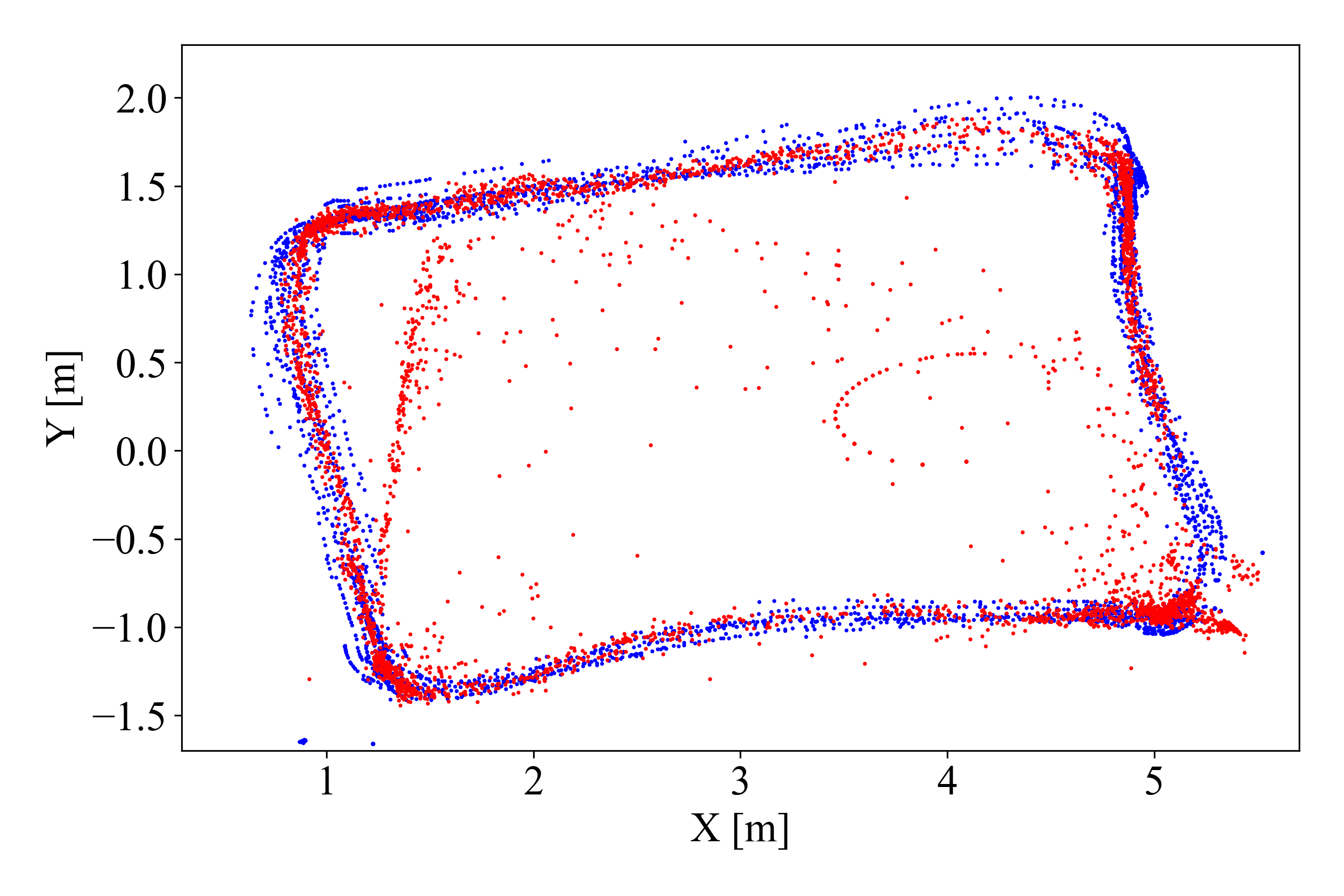}
        \subcaption{RNN-Decay}\label{fig:estimate_rnn_decay}
    \end{minipage} \\
    \begin{minipage}[t]{0.475\hsize}
        \centering
        \includegraphics[width=1\hsize]{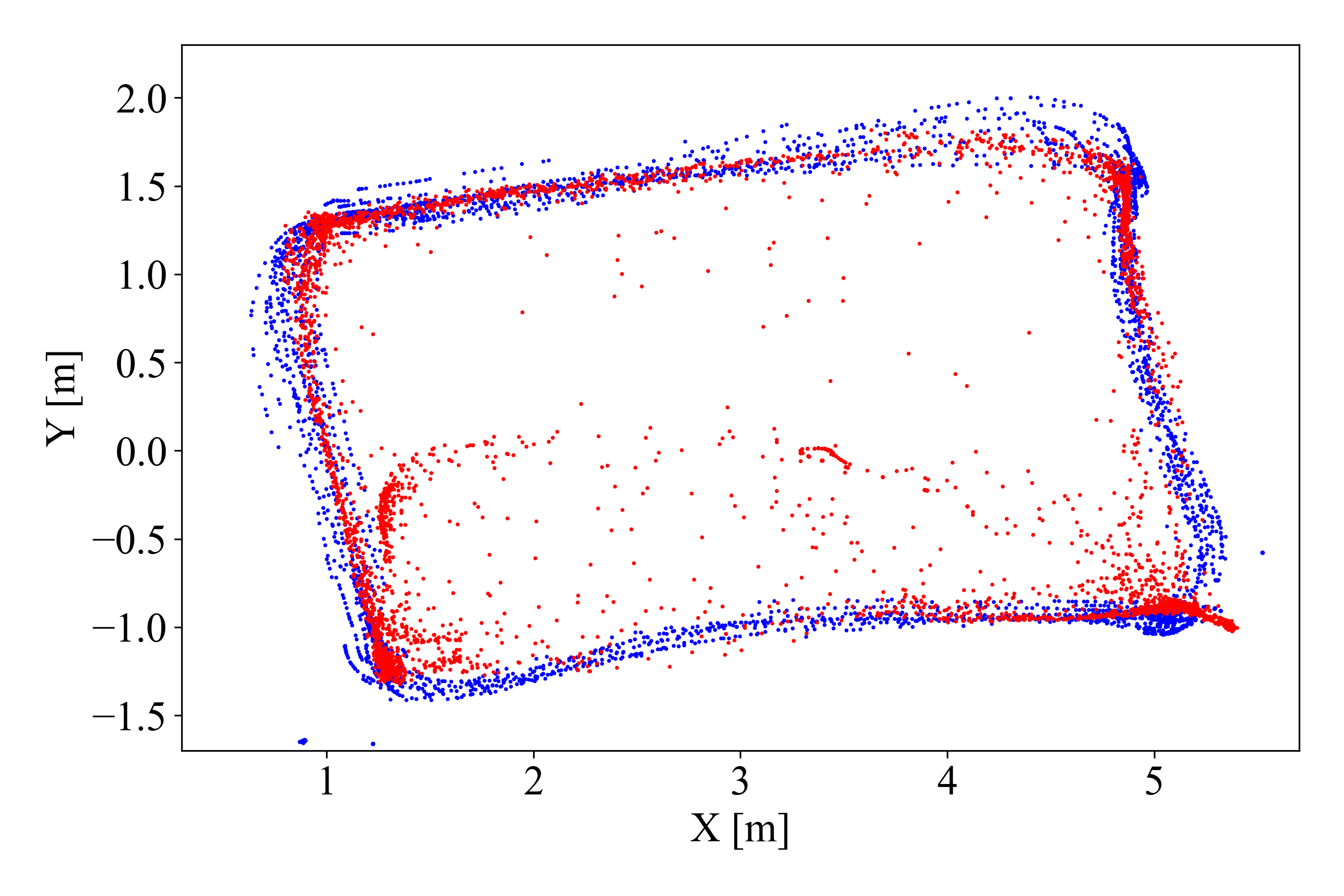}
        \subcaption{RNN-$\Delta$}\label{fig:estimate_rnn_delta}
    \end{minipage}
    \begin{minipage}[t]{0.475\hsize}
        \centering
        \includegraphics[width=1\hsize]{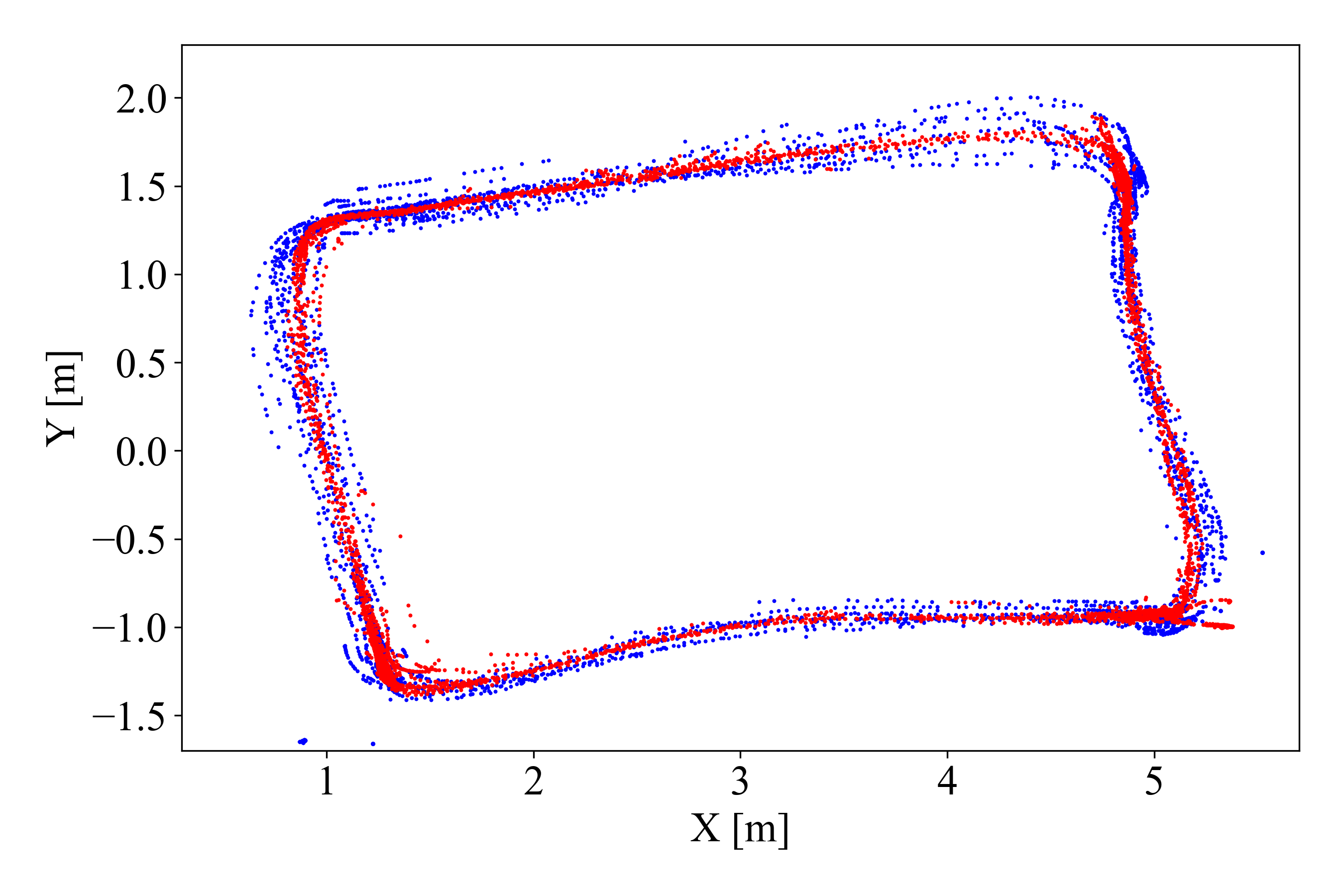}
        \subcaption{NDF}\label{fig:estimate_ndf}
    \end{minipage}
    \caption{Visualization of estimated trajectories (red) and groundtruth (blue) for selected multi-band baseline methods and the proposed NDF.}
    \label{fig:performance_visualization_main}
\end{figure}

\subsection{Impact of Sequence Length \texorpdfstring{$\Delta T_w$}{}}
In the following, we investigate the impact of sequence length $\Delta T_w$ on the trajectory estimation performance.
Given the frame rate of about $5$ Hz for CSI and about $1$ Hz for beam SNR, the number of effective samples is proportional to the sequence length $\Delta T_w$.
For a given sequence length $\Delta T_w$, we follow the random split protocol to segment the raw data into non-overlapping $\Delta T_w$-sec sequences with $\Delta T_w = \{2, 8\}$ seconds.

Table~\ref{tab:eval_window_size} lists the trajectory estimation errors in terms of mean, median, and the $90$-th percentile of the CDF for three choices of $\Delta T_w$.
Overall, it confirms that, the longer the sequence, the better the trajectory estimation performance.
In the case of $\Delta T_w=2$ seconds, there might not be sufficient beam SNR samples for latent dynamic learning as the frame rate is limited to about $1$ Hz.
The choice of $\Delta T_w=8$ seconds appears to give lower median and CDF@0.9 localization errors while keeping the mean error close to that of $\Delta T_w=5$ seconds.

\begin{table}[t]
    \centering
    \caption{Impact of sequence length $\Delta T_w$. }
    \label{tab:eval_window_size}
    \begin{tabular}{cccc} \toprule
        Sequence length & Mean & Median & CDF@0.9 \\ \midrule
         2 seconds & 0.481 & 0.233 & 1.20\\
         5 seconds & \textbf{0.263} & 0.148 & 0.611 \\ 
         8 seconds & 0.270 & \textbf{0.136} & \textbf{0.547} \\ \bottomrule
    \end{tabular}
\end{table}

\subsection{Impact of Fusion Scheme}
\begin{table}[t]
    \centering
    \caption{The impact of the three fusion schemes in Sec.~\ref{sec:fusion_scheme} on the trajectory estimation error (m). }
    \begin{tabular}{cccc} \toprule
         \textbf{Fusion scheme} & Mean & Median & CDF@0.9 \\ \midrule
         MLP & \textbf{0.263} & \textbf{0.148} & \textbf{0.611} \\
         Pairwise Interaction & 0.397 & 0.230 & 0.903 \\
         Weighted Importancee & 0.287 & 0.164 & 0.618 \\ \bottomrule
    \end{tabular}
    \label{tab:fusion_scheme}
\end{table}
In the following, we examine the impact of the three fusion schemes in Sec.~\ref{sec:fusion_scheme} on trajectory estimation accuracy, using the random split protocol of nonoverlapping $5$-sec input sequences. 
As shown in Table~\ref{tab:fusion_scheme}, the MLP fusion scheme delivers the best results in terms of mean, median, and 90th-percentile CDF, with the weighted importance fusion scheme exhibiting nearly identical performance.

Moreover, the three fusion schemes outperform all multi-band baseline methods listed in Table~\ref{tab:error_statistics}. 
This seems to imply that once the latent states between the beam SNR and CSI are aligned, the choice of fusion scheme has only a marginal impact on the final localization performance.

\subsection{Generalization under Temporal Split}
\label{sec:extraplo}
\begin{table}[t]
    \centering
    \caption{ Localization error (m) under temporal split as a function of training data ratio $s\%$.}
    \label{tab:eval_training_ratio}
    \begin{tabular}{cccc} \toprule
         Training Data Ratio & Mean & Median & CDF@0.9 \\ \midrule
         20\% & 0.647 & 0.250 & 2.07 \\
         40\% & 0.749 & 0.183 & 3.81 \\ 
         60\% & \textbf{0.454} & \textbf{0.169} & \textbf{0.981} \\
         80\% & 0.501 & 0.183 & 1.17 \\ \bottomrule
    \end{tabular}
\end{table}

Under \textbf{temporal split}, sequences in the training and test sets are distinctly separated in the temporal domain such that they do not intertwine. This separation allows us to effectively assess the generalization capability on future Wi-Fi measurements. We consider various training data ratios of $s\%={20\%, 40\%, 60\%, 80\%}$ corresponding to, respectively, $(20:80, 40:60, 60:40, 80:20)$ training-test data split ratios. 
For better use of training data when the training data size is small, we segment the training data into $5$-sec overlapping sequences using a stepsize of $1$ second.

Table~\ref{tab:eval_training_ratio} shows the localization errors for various choices of $s$. It is seen that a temporal training-test split ratio of $60:40$ provides the best localization performance across all evaluated metrics. The NDF with a temporal training-test split ratio of $80:20$ results in less accurate performance compared to the random training-test split ratio of $80:10:10$ in Table~\ref{tab:error_statistics}. For example, the mean localization error increases from $26.3$ cm under the random split to $50.1$ cm under the temporal split.  
Such degradation is anticipated due to the temporal fluctuation in Wi-Fi measurements, which may stem from channel instability over time.

\begin{figure*}[t]
    \centering
    \begin{minipage}[b]{0.19\hsize}
        \centering
        \includegraphics[width=1\hsize]{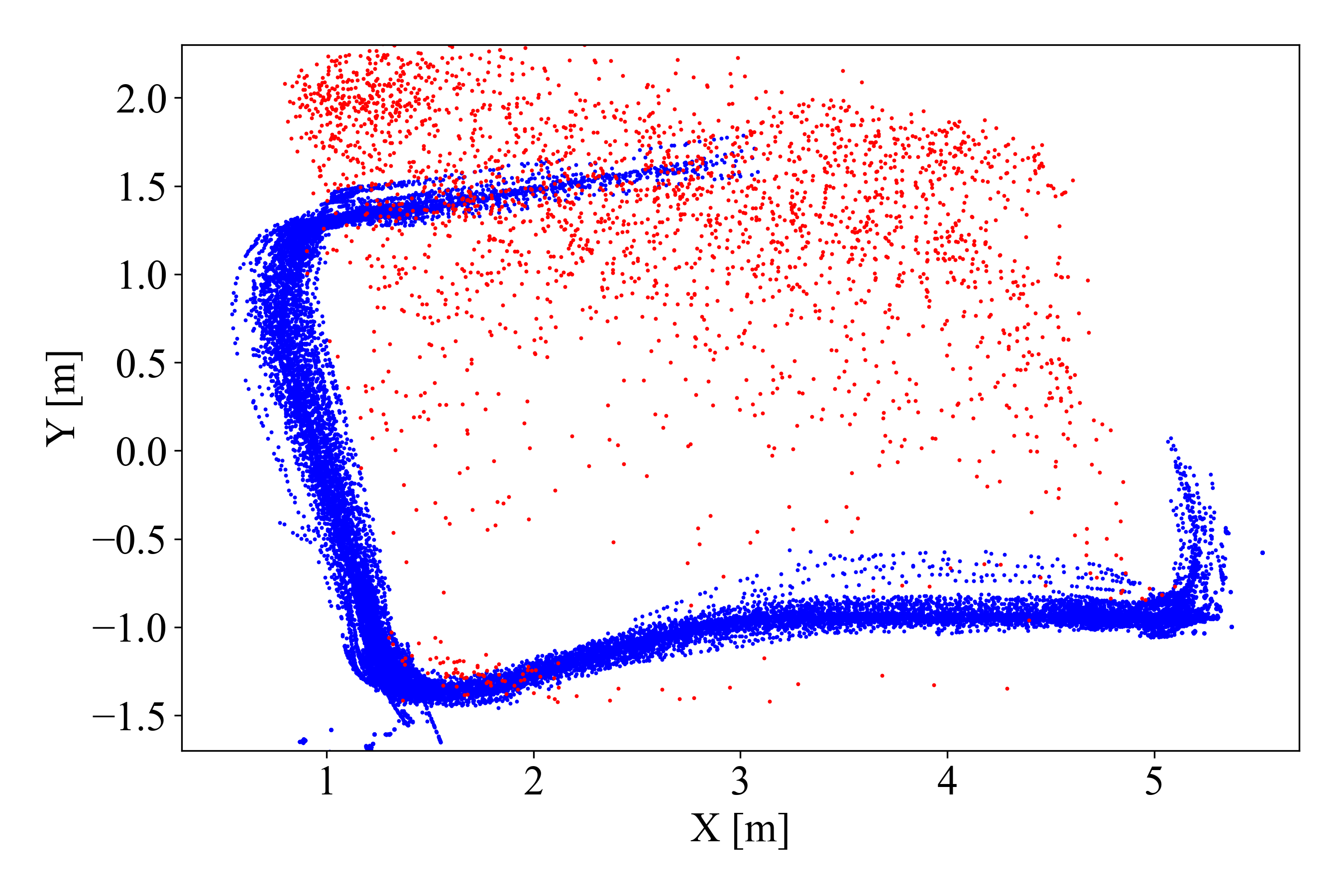}
        \subcaption{LinearInt fusion}\label{fig:corner_crop_linear}
    \end{minipage}
    \begin{minipage}[b]{0.19\hsize}
        \centering
        \includegraphics[width=1\hsize]{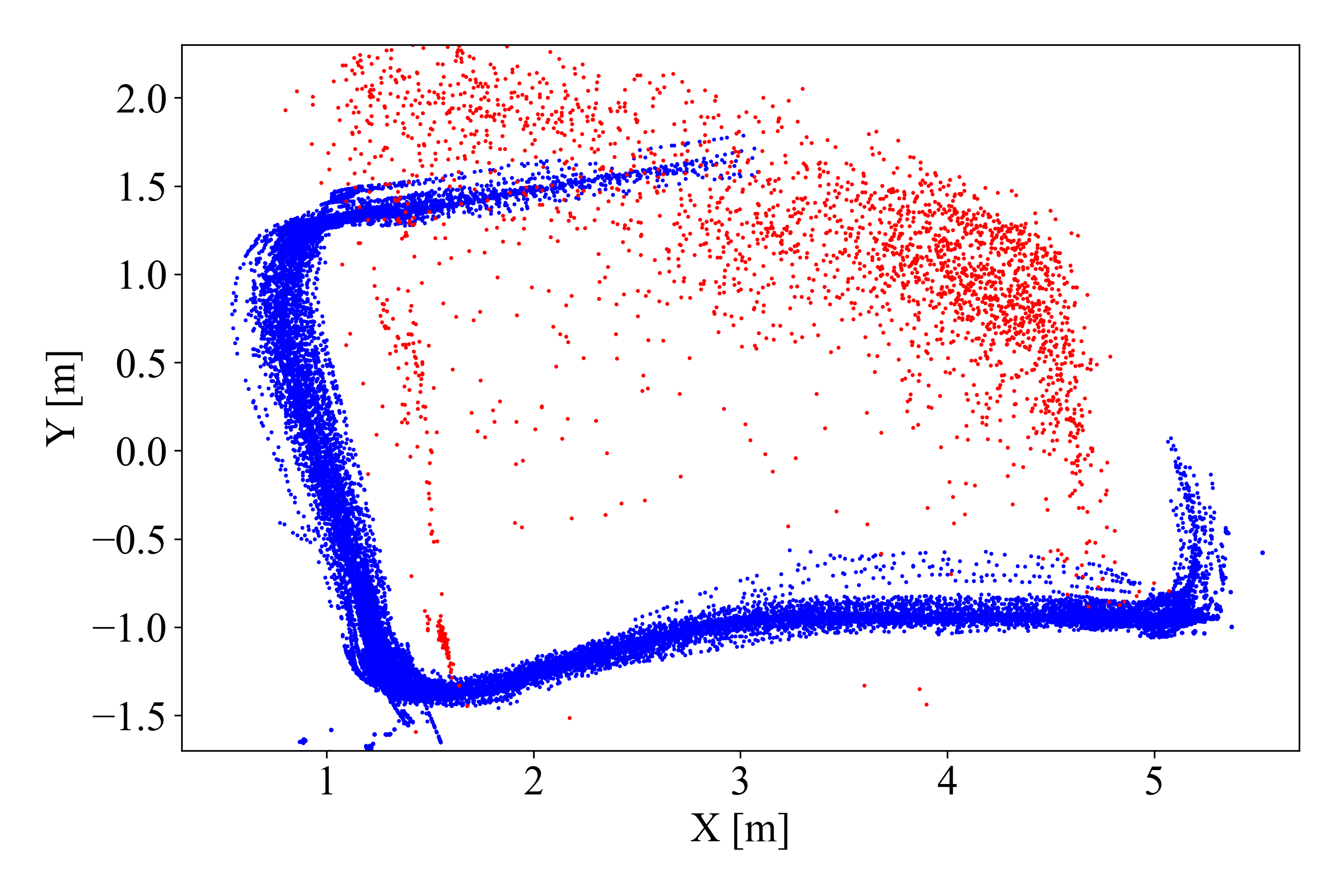}
        \subcaption{NearestInt fusion}\label{fig:corner_crop_nearest}
    \end{minipage}
    \begin{minipage}[b]{0.19\hsize}
        \centering
        \includegraphics[width=1\hsize]{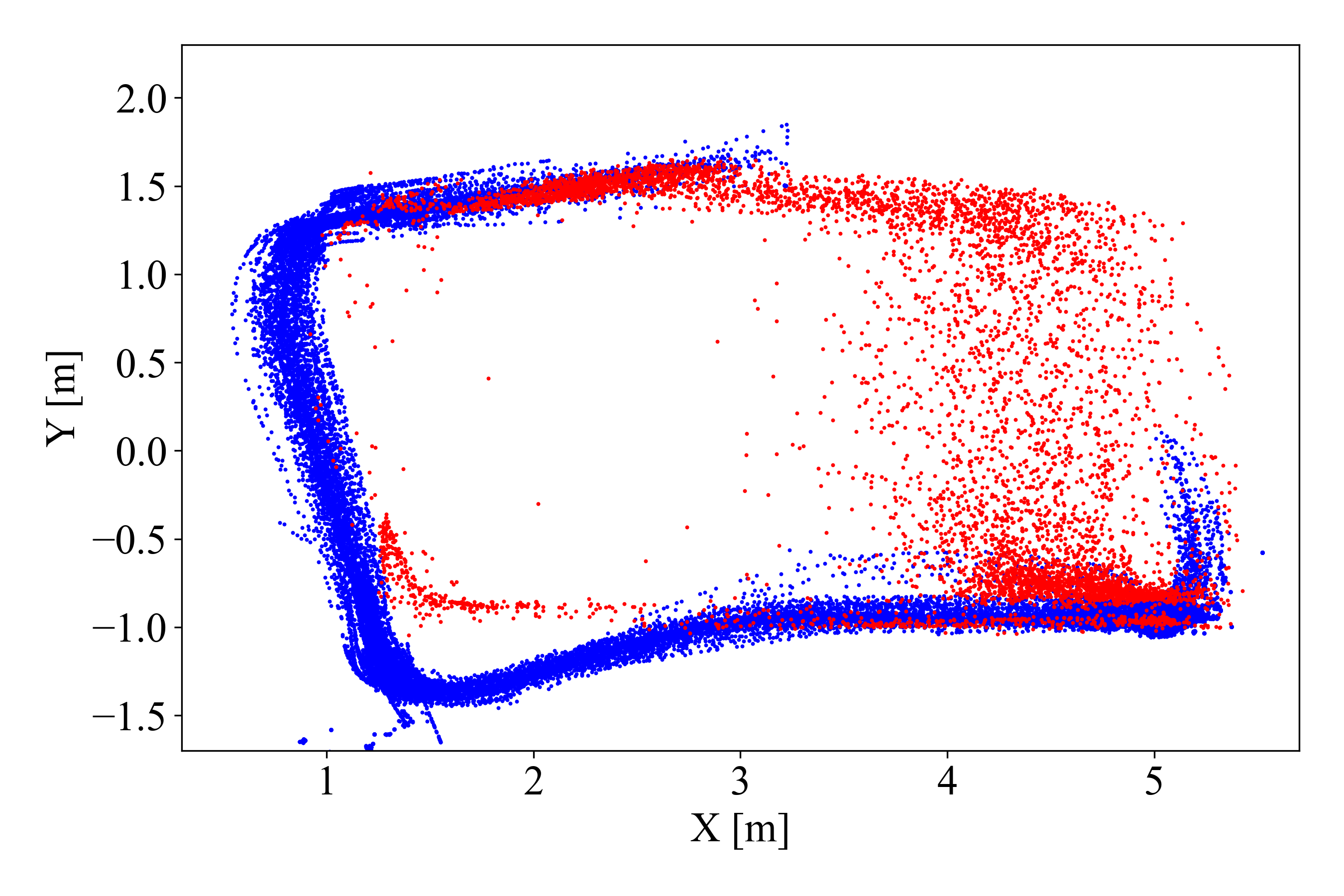}
        \subcaption{RNN-Decay fusion}\label{fig:corner_crop_rnn_decay}
    \end{minipage}
    \begin{minipage}[b]{0.19\hsize}
        \centering
        \includegraphics[width=1\hsize]{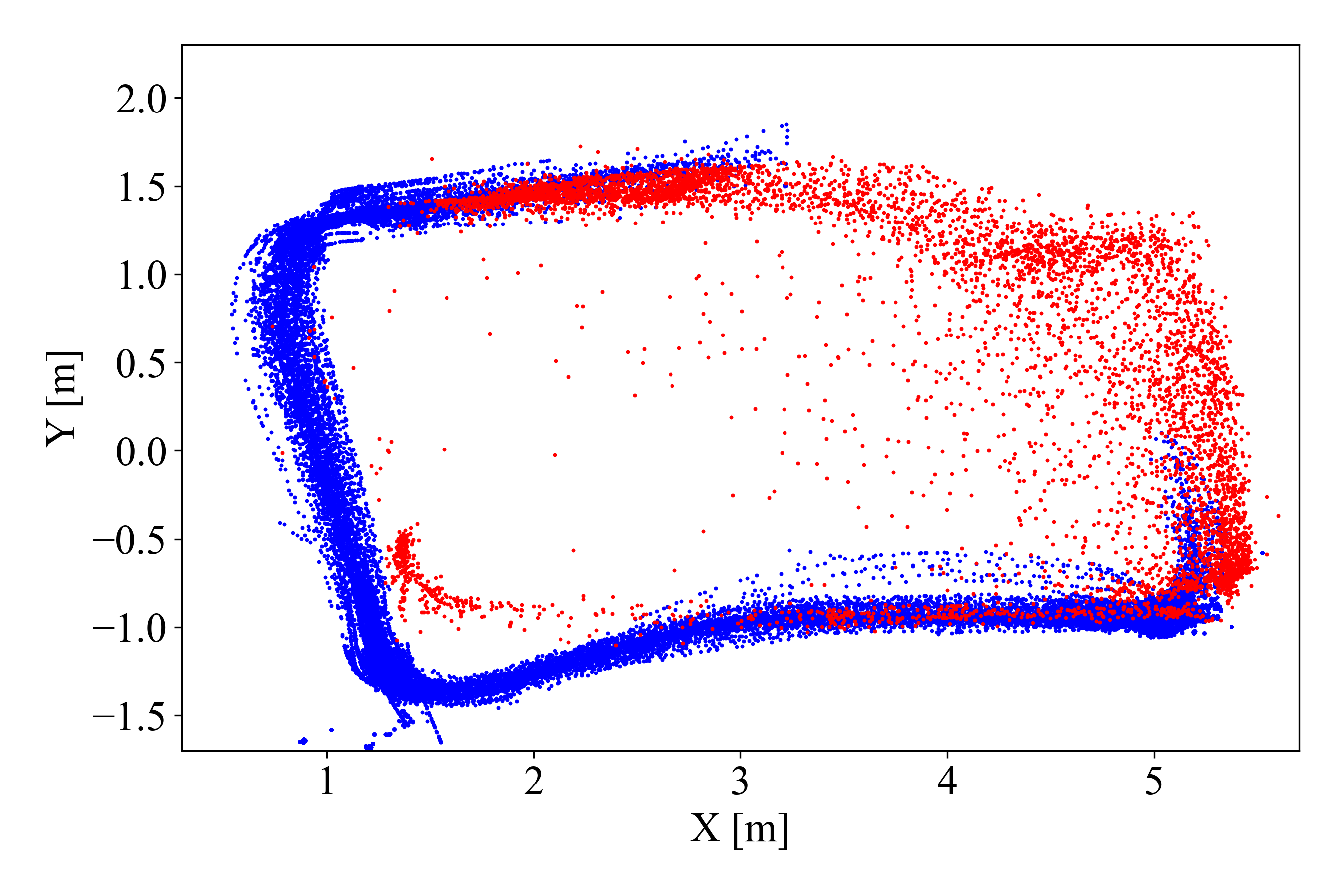}
        \subcaption{RNN-$\Delta$ fusion}\label{fig:corner_crop_rnn_delta}
    \end{minipage}
    \begin{minipage}[b]{0.19\hsize}
        \centering
        \includegraphics[width=1\hsize]{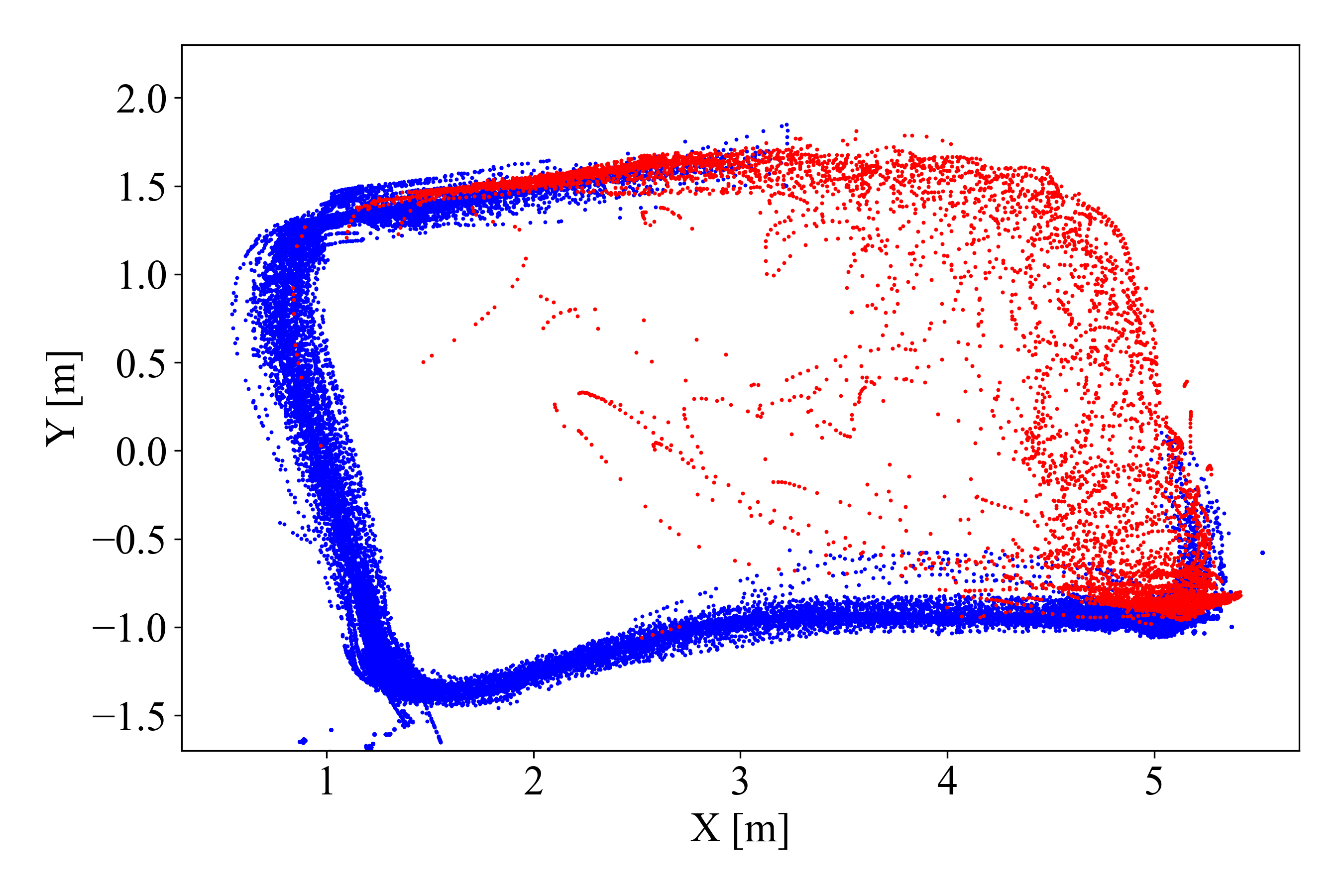}
        \subcaption{NDF}\label{fig:corner_crop_ndf}
    \end{minipage}
    \caption{Generalization capability to unseen locations (upper right corner) is illustrated with training trajectories in blue. Ideally, the estimated trajectories in red should complement the blue training trajectories to form a complete rectangular path. }
    \label{fig:corner_crop}
\end{figure*}

\begin{figure*}[t]
    \centering
    \begin{minipage}[t]{0.32\hsize}
        \includegraphics[width=1\hsize]{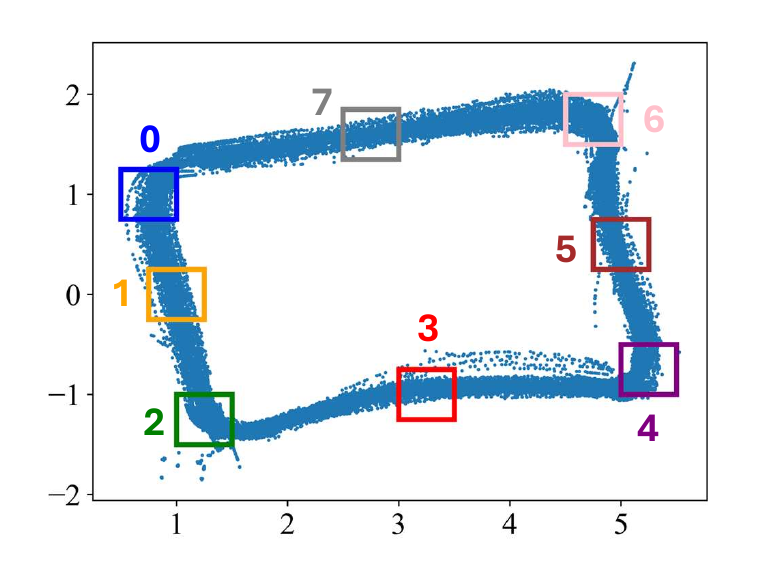}
        \subcaption{Selected local regions for latent space inspection.}\label{fig:eight_areas}
    \end{minipage}
    \begin{minipage}[t]{0.65\hsize}
        \includegraphics[width=1\hsize]{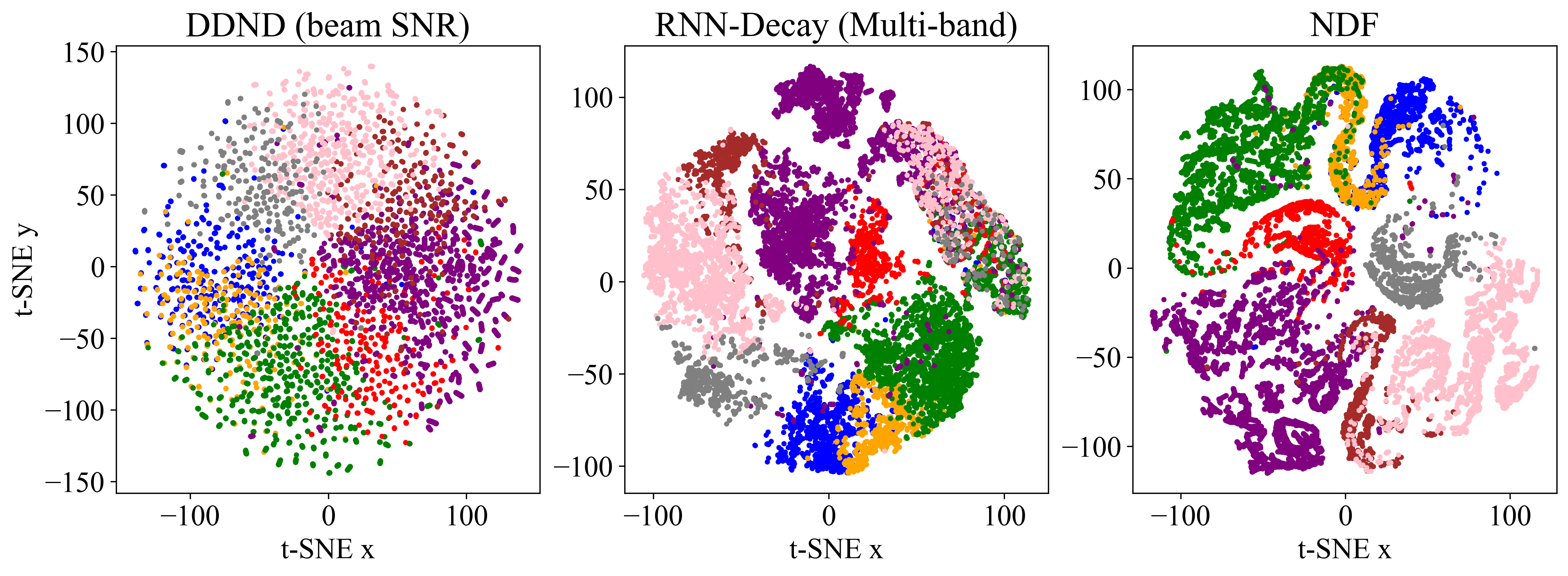}
        \subcaption{Latent states from selected  regions projected on 2D plane by t-SNE.}\label{fig:latent_tsne_compare}
    \end{minipage}
    \caption{Visualization of learned latent states from selected local regions, with matching colors between the local regions and their corresponding latent states.}
    \label{fig:latenet_tsne}
\end{figure*}

\subsection{Generalization under Coordinate Split}
\label{sec:unseen}

Under \textbf{coordinate split}, we can test the generalization capability on test data collected from unseen positions. Fig.~\ref{fig:corner_crop} illustrates the estimated trajectories (in red dots) alongside the ground truth trajectories in the upper right corner, with training trajectories plotted in blue. As shown in Fig.~\ref{fig:corner_crop} (a) and (b), the frame-to-frame fusion baseline methods (LinearInt and NearestInt) fail to leverage latent dynamics, resulting in scattered estimated trajectories. In contrast, the sequence-to-sequence baseline methods, specifically RNN-Decay fusion and RNN-$\Delta$ fusion in Fig.~\ref{fig:corner_crop} (c) and (d), demonstrate improved alignment between the estimated red trajectories and the training blue trajectories. Fig.~\ref{fig:corner_crop} (e) presents the NDF results, clearly showing that the estimated trajectories effectively complement the training trajectories, closely mirroring the ground truth rectangular trajectories.

\subsection{Latent Space Visualization}
As shown in Fig.~\ref{fig:latenet_tsne} (a), 
we identify $8$ local regions along the trajectory. We gather all data frames with their corresponding ground truth locations from the same region and map them into latent states using selected baseline methods and the proposed NDF method.  These high-dimensional latent states are then visualized by projecting them onto a 2D plane using t-distributed Stochastic Neighbor Embedding (t-SNE). 

Fig.~\ref{fig:latenet_tsne} (b) presents the t-SNE visualization results for the single-band baseline (beam SNR-based DDND), the RNN-Decay fusion baseline, and the proposed NDF method. The sequence-to-sequence RNN-Decay fusion baseline exhibits much clearer separation compared to  the single-band beam SNR-based DDND, highlighting the advantages of utilizing multi-band Wi-Fi channel measurements in our experiment. Nonetheless, it is seen that the latent states from regions $2$ and $6$ overlap within the upper right cluster. 

In contrast, Fig.~\ref{fig:latent_tsne_compare} demonstrates that our NDF learns a compact and well-separated representation in the latent space, with denser latent distributions for each region. These results further suggest that the low-dimensional latent space effectively preserves the trajectory geometry within the NDF framework. Notably, each latent cluster is spatially connected to the edge of its adjacent cluster, creating a continuous latent space that seamlessly transitions from one region to another.

\section{Conclusion}
\label{sec:conclusion}
In this paper, we introduce the NDF framework, which utilizes asynchronous multi-band Wi-Fi channel measurements to estimate trajectories in a continuous-time manner. This is achieved through a multiple-encoder, multiple-decoder architecture that aligns latent states across different input sequences and fuses them for trajectory estimation. Latent state alignment is facilitated by a learnable ODE model and the initial latent conditions from the encoders.  Evaluated with real-world multi-band Wi-Fi data, the NDF framework demonstrates significant performance enhancements compared with a comprehensive set of single-band and multi-band baseline methods.

\bibliographystyle{IEEEbib}
\bibliography{bib/bib_localization}

\appendices
\section{CSI Calbiration and Embedding}
\label{sec:csiEmbedding}
The extracted CSI suffers from  both magnitude and phase offsets including carrier frequency offset~(CFO), sample time offset~(STO) \cite{kotaru2015spotfi,  guo2017wifi, wang2017phasebeat, zeng2018fullbreathe, zhang2022practical, zhang2020calibrating}. We choose the SpotFi~\cite{kotaru2015spotfi} calibration method to eliminate the linear phase offset caused by STO and CSI conjugate multiplication to cancel out packet-wise random phase offset and improve the stability of the waveform. 
The calibration is performed packet-wise and antenna-wise at the receiving side.
For each receiving antenna, we first unwrap the CSI phase, then we obtain the best linear fit of the unwrapped phase as
\begin{equation}
    [\hat{\tau}_{i, n}, \hat{\beta}_{i,n}] = \arg\min_{\tau, \beta} \sum_{j, k=1}^{N_{\mathrm Rx}, N_{\mathrm s}}  (\psi_n(i, j, k) - 2\pi f_\delta (k-1) \tau + \beta)^2,
\end{equation}
where $\psi_n(i, j, k) = \angle {C}_n(i, j, k)$ is the unwrapped phase of the $n$-th packet  from transmitting antenna $i$ and receiving antenna $j$ at subcarrier $k$, and $f_\delta$ is the frequency spacing between two adjacent subcarriers.
We obtain the calibrated phase by subtracting phase offset as
\begin{equation}
    \hat{\psi}_n(i, j, k) = \psi_n(i, j, k) - 2\pi f_\delta (k-1) \hat{\tau}_{i,n}.
\end{equation}
We further perform CSI conjugate multiplication across the receiving antennas to remove random phase fluctuation over packets~\cite{zeng2018fullbreathe}.
This leads to the calibrated CSI element
\begin{equation}
    \tilde{C}_n(i, j, k) = C_n(i, j, k){C^{*}_n(i, j+1, k)},
\end{equation}
where $j=1, \cdots, N_{\mathrm Rx}-1$, and $*$ represents the complex conjugate.
Grouping all calibrated CSI elements $\tilde{C}_n(i, j, k)$ over transmitting antenna $i$, receiving antenna $j$, and subcarrier $k$, the calibrated CSI tensor is given by $\tilde{\Cbf}_n \in \mathbb{C}^{N_{\mathrm Tx} \times (N_{\mathrm Rx}-1) \times N_{\mathrm s}}$.

To balance between the two input (CSI and beam SNR) sequences, we employ a pretrained convolutional autoencoder (CAE) to compresscalibrated CSI tensor $\tilde{\Cbf}_n$ into an embedding vector $\cbf_n \in \mathbb{R}^{M_c \times 1}$ with the following steps:
\begin{itemize}
\item \textbf{Complex-to-Real Conversion}: we convert the complex-valued $\tilde{\Cbf}_n$ into a real-valued matrix $\Cbf^f_n \in \mathbb{R}^{N_{\mathrm Tx}(N_{\mathrm Rx}-1)N_s \times 4}$. This is achieved by by splitting each element into four parts: real, imaginary, phase, and magnitude. Each of these parts is then vectorized into a 1D vector and putting these four vectors togther yields $\Cbf^f_n$. 
\item \textbf{Embedding from Autoencoder}: 
The real-valued matrix $\Cbf^f_n$ is fed to the CAE as $\cbf_n =\mathcal{E}_{\thetabf_{\mathrm AE}^e}(\Cbf_n^f)$ and $\hat{\Cbf}_n^f = \mathcal{D}_{\thetabf_{\mathrm AE}^d}(\cbf_n)$, 
where $\cbf_n$ is the CSI embedding vector, $\mathcal{E}_{\thetabf_{\mathrm AE}^e}$ and $\mathcal{D}_{\thetabf_{\mathrm AE}^d}$ represent the encoder and decoder of CAE, respectively, and $\hat{\Cbf}_n^f$ is the reconstructed real-valued CSI matrix at the decoder output. 
The CAE is pretrained by minimizing the reconstruction error between $\Cbf^f_n$ and $\hat{\Cbf}_n^f$.
\end{itemize}

\section{LSTM Update Step}
\label{RNN-LSTM}
Given the measurement $\sbf_n$ at time step $n$ and the auxiliary variable $\tilde{\hbf}_n$, one can use a standard LSTM unit to update the latent variable $\hbf_{n} =\mathcal{R}(\tilde{\hbf}_n, {\sbf}_n; \thetabf), \quad n=0, 1, \cdots, N$,
where $\mathcal{R}(\cdot, \cdot |\thetabf)$ is implemented with the following process (with abuse of notation)
\begin{align}
\tilde{\cbf}_n &= \tanh \left(\Wbf_{r c} {\sbf}_{n}+\Wbf_{h c} \tilde{\hbf}_n+\bbf_{c}\right),  \label{candidate} \\
\fbf_{n} &=\sigma\left(\Wbf_{rf} {\sbf}_{n}+\Wbf_{h f} \tilde{\hbf}_n+\bbf_{f}\right), \label{forget} \\
\ibf_{n} &=\sigma \left(\Wbf_{ri} {\sbf}_{n}+\Wbf_{hi} \tilde{\hbf}_n+\bbf_{i}\right). \label{input}
\end{align}
The above process consists of three \emph{gates}: 
\begin{itemize}
    \item a memory gate of \eqref{candidate} uses the tanh function to combine the auxiliary hidden state $\tilde{\hbf}_n$ and the current input $\sbf_n$ into a value range of $(-1, 1)$. 
    \item a forget gate of \eqref{forget} also acts on $(\tilde{\hbf}_n, \sbf_n)$ but compresses the value into $(0, 1)$ with the sigmoid function $\sigma(\cdot)$ to determine how much of the old memory should retain. 
    \item an input gate of \eqref{input} compresses $(\tilde{\hbf}_n,\sbf_n)$ into another value in between $0$ and $1$ and decides how much information we should take from the new input $\sbf_n$,
\end{itemize}
along with weight matrices $\Wbf_{rc/rf/ri/hc/hf/hi}$ and bias terms $\bbf_{c/f/i}$.  Then new hidden state $\hbf_n$ is updated as
$\hbf_{n} =\tanh \left(\hat{\cbf}_{n}\right)  \odot  \obf_{n}$, 
where the new memory variable $\hat{\cbf}_n$ updates its ``old" memory $\hat{\cbf}_{n-1}$ passing through the ``current" forget gate output $\fbf_n$ and adds new memory cell $\tilde{\cbf}_n$ weighted by the ``current" input gate output $\ibf_n$: 
$\hat{\cbf}_{n} =\fbf_{n} \odot \hat{\cbf}_{n-1}+\ibf_{n} \odot \tilde{\cbf}_n$, 
and the output gate $\obf_n$ is computed as
\begin{align}
\obf_{n} &=\sigma\left(\Wbf_{r o} {\sbf}_{n}+\Wbf_{h o} \tilde{\hbf}_n+\Wbf_{c o} \odot \hat{\cbf}_{n}+ \bbf_{o}\right). \label{output}
\end{align}
It is seen that the parameters $\thetabf$ in the LSTM update step is given as $\thetabf=\{\Wbf_{rc/rf/ri/hc/hf/hi/ro/ho/co}, \bbf_{c/f/i/o} \}$. 

\section{General derivation of ELBO}\label{appendix_elbo}
Evidence lower bound, or ELBO, is a lower bound on the log-likelihood of observed data.
We first express the log-likelihood of the input data $\xbf$ as
\small
\begin{equation}\label{elboStart}
    \begin{split}
        \log p(\xbf) &= \log p(\xbf) \int q(\zbf|\xbf) d\zbf \\
        &= \int q(\zbf|\xbf) \left(\log \dfrac{q(\zbf|\xbf)}{p(\zbf|\xbf)} + \log \dfrac{p(\xbf, \zbf)}{q(\zbf|\xbf)}\right) d\zbf \\
        &= D_{\rm KL}[q(\zbf|\xbf)||p(\zbf|\xbf)] + \int q(\zbf|\xbf) \log \dfrac{p(\xbf, \zbf)}{q(\zbf|\xbf)} d\zbf, \\
    \end{split}
\end{equation}
where $D_{\rm KL}[\cdot || \cdot]$ is the {Kullback–Leibler} (KL) divergence between two given distributions.
Given that $D_{\rm KL}[q(\zbf|\xbf)||p(\zbf|\xbf)] \geq 0$, \eqref{elboStart} follows as \cite{kingma2013auto}
\begin{align}\label{elbo_iniquality}
    \log p(\xbf) \geq & 
 \int q(\zbf|\xbf) \log q(\xbf|\zbf) d\zbf + \int q(\zbf|\xbf) \log \dfrac{p(\zbf)}{q(\zbf|\xbf)} d\zbf \notag \\
        &= \mathbb{E}_{q(\zbf|\xbf)}\left[\log q(\xbf|\zbf)\right] - D_{\rm KL}[q(\zbf|\xbf)||p(\zbf)]. 
\end{align}

We extend the above lower bound to the case where two inputs $\xbf$ and $\ybf$ with corresponding $\zbf_x$ and $\zbf_y$ as
\begin{equation}
    \mathbb{E}_{q(\zbf_x, \zbf_y|\xbf, \ybf)}\left[\log q(\xbf, \ybf|\zbf_x, \zbf_y)\right] - D_{\rm KL}[q(\zbf_x, \zbf_y|\xbf, \ybf)||p(\zbf_x, \zbf_y)]. \notag 
\end{equation}
The KL divergence term can be decomposed to the sum of two KL divergence terms by using the independent assumptions between $\xbf$ and $\ybf$ and between $\zbf_x$ and $\zbf_y$,
\small
\begin{align} \label{KLindep}
    &D_{\rm KL}[q(\zbf_x, \zbf_y|\xbf, \ybf)||p(\zbf_x, \zbf_y)] \notag \\
    &= \int \int q(\zbf_x|\xbf)q(\zbf_y|\ybf) \left(\log \dfrac{q(\zbf_x|\xbf)}{p(\zbf_x)} + \log \dfrac{q(\zbf_y|\ybf)}{p(\zbf_y)}\right) d\zbf_x d\zbf_y \notag \\
    &= \underbrace{\int q(\zbf_y|\ybf) d\zbf_y}_{=1} \int q(\zbf_x|\xbf) \log \dfrac{q(\zbf_x|\xbf)}{p(\zbf_x)}d\zbf_x \notag \\
    &\quad +  \underbrace{\int q(\zbf_x|\xbf) d\zbf_x}_{=1} \int q(\zbf_y|\ybf) \log \dfrac{q(\zbf_y|\ybf)}{p(\zbf_y)} d\zbf_y \notag \\
    &= D_\mathrm{KL}[q(\zbf_x|\xbf) || p(\zbf_x)] + D_\mathrm{KL}[q(\zbf_y|\ybf) || p(\zbf_y)].
\end{align}

\end{document}

%% file: utils/com.tex

\newcommand{\ba}{\begin{array}}
\newcommand{\ea}{\end{array}}
\newcommand{\be}{\begin{displaymath}}
\newcommand{\ee}{\end{displaymath}}
\newcommand{\ben}{\begin{equation}}
\newcommand{\een}{\end{equation}}
\newcommand{\bena}{\begin{eqnarray}}
\newcommand{\eena}{\end{eqnarray}}
\newcommand{\beqa}{\begin{eqnarray*}}
\newcommand{\enqa}{\end{eqnarray*}}
\newcommand{\f}{\frac}
\newcommand{\bc}{\begin{center}}
\newcommand{\ec}{\end{center}}
\newcommand{\bi}{\begin{itemize}}
\newcommand{\ei}{\end{itemize}}
\newcommand{\benu}{\begin{enumerate}}
\newcommand{\eenu}{\end{enumerate}}
\newcommand{\bdes}{\begin{description}}
\newcommand{\edes}{\end{description}}
\newcommand{\bt}{\begin{tabular}}
\newcommand{\et}{\end{tabular}}
\newcommand{\vs}{\vspace}
\newcommand{\hs}{\hspace}

\newcommand \thetabf{{\mbox{\boldmath$\theta$\unboldmath}}}
\newcommand{\Phibf}{\mbox{${\bf \Phi}$}}
\newcommand{\Psibf}{\mbox{${\bf \Psi}$}}
\newcommand \alphabf{\mbox{\boldmath$\alpha$\unboldmath}}
\newcommand \betabf{\mbox{\boldmath$\beta$\unboldmath}}
\newcommand \gammabf{\mbox{\boldmath$\gamma$\unboldmath}}
\newcommand \deltabf{\mbox{\boldmath$\delta$\unboldmath}}
\newcommand \epsilonbf{\mbox{\boldmath$\epsilon$\unboldmath}}
\newcommand \zetabf{\mbox{\boldmath$\zeta$\unboldmath}}
\newcommand \etabf{\mbox{\boldmath$\eta$\unboldmath}}
\newcommand \iotabf{\mbox{\boldmath$\iota$\unboldmath}}
\newcommand \kappabf{\mbox{\boldmath$\kappa$\unboldmath}}
\newcommand \lambdabf{\mbox{\boldmath$\lambda$\unboldmath}}
\newcommand \mubf{\mbox{\boldmath$\mu$\unboldmath}}
\newcommand \nubf{\mbox{\boldmath$\nu$\unboldmath}}
\newcommand \xibf{\mbox{\boldmath$\xi$\unboldmath}}
\newcommand \pibf{\mbox{\boldmath$\pi$\unboldmath}}
\newcommand \rhobf{\mbox{\boldmath$\rho$\unboldmath}}
\newcommand \sigmabf{\mbox{\boldmath$\sigma$\unboldmath}}
\newcommand \taubf{\mbox{\boldmath$\tau$\unboldmath}}
\newcommand \upsilonbf{\mbox{\boldmath$\upsilon$\unboldmath}}
\newcommand \phibf{\mbox{\boldmath$\phi$\unboldmath}}
\newcommand \varphibf{\mbox{\boldmath$\varphi$\unboldmath}}
\newcommand \chibf{\mbox{\boldmath$\chi$\unboldmath}}
\newcommand \psibf{\mbox{\boldmath$\psi$\unboldmath}}
\newcommand \omegabf{\mbox{\boldmath$\omega$\unboldmath}}
\newcommand \Sigmabf{\hbox{$\bf \Sigma$}}
\newcommand \Upsilonbf{\hbox{$\bf \Upsilon$}}
\newcommand \Omegabf{\hbox{$\bf \Omega$}}
\newcommand \Deltabf{\hbox{$\bf \Delta$}}
\newcommand \Gammabf{\hbox{$\bf \Gamma$}}
\newcommand \Thetabf{\hbox{$\bf \Theta$}}
\newcommand \Lambdabf{\hbox{$\bf \Lambda$}}
\newcommand \Xibf{\hbox{\bf$\Xi$}}
\newcommand \Pibf{\hbox{\bf$\Pi$}}

\newcommand \abf{{\bf a}}
\newcommand \bbf{{\bf b}}
\newcommand \cbf{{\bf c}}
\newcommand \dbf{{\bf d}}
\newcommand \ebf{{\bf e}}
\newcommand \fbf{{\bf f}}
\newcommand \gbf{{\bf g}}
\newcommand \hbf{{\bf h}}
\newcommand \ibf{{\bf i}}
\newcommand \jbf{{\bf j}}
\newcommand \kbf{{\bf k}}
\newcommand \lbf{{\bf l}}
\newcommand \mbf{{\bf m}}
\newcommand \nbf{{\bf n}}
\newcommand \obf{{\bf o}}
\newcommand \pbf{{\bf p}}
\newcommand \qbf{{\bf q}}
\newcommand \rbf{{\bf r}}
\newcommand \sbf{{\bf s}}
\newcommand \tbf{{\bf t}}
\newcommand \ubf{{\bf u}}
\newcommand \vbf{{\bf v}}
\newcommand \wbf{{\bf w}}
\newcommand \xbf{{\bf x}}
\newcommand \ybf{{\bf y}}
\newcommand \zbf{{\bf z}}
\newcommand \rbfa{{\bf r}}
\newcommand \xbfa{{\bf x}}
\newcommand \ybfa{{\bf y}}
\newcommand \Abf{{\bf A}}
\newcommand \Bbf{{\bf B}}
\newcommand \Cbf{{\bf C}}
\newcommand \Dbf{{\bf D}}
\newcommand \Ebf{{\bf E}}
\newcommand \Fbf{{\bf F}}
\newcommand \Gbf{{\bf G}}
\newcommand \Hbf{{\bf H}}
\newcommand \Ibf{{\bf I}}
\newcommand \Jbf{{\bf J}}
\newcommand \Kbf{{\bf K}}
\newcommand \Lbf{{\bf L}}
\newcommand \Mbf{{\bf M}}
\newcommand \Nbf{{\bf N}}
\newcommand \Obf{{\bf O}}
\newcommand \Pbf{{\bf P}}
\newcommand \Qbf{{\bf Q}}
\newcommand \Rbf{{\bf R}}
\newcommand \Sbf{{\bf S}}
\newcommand \Tbf{{\bf T}}
\newcommand \Ubf{{\bf U}}
\newcommand \Vbf{{\bf V}}
\newcommand \Wbf{{\bf W}}
\newcommand \Xbf{{\bf X}}
\newcommand \Ybf{{\bf Y}}
\newcommand \Zbf{{\bf Z}}
\newcommand \Rssbf{{\bf R_{ss}}}
\newcommand \Ryybf{{\bf R_{yy}}}
\newcommand{\otheta}{\stackrel{\circ}{\theta}}
\newcommand{\defeq}{\stackrel{\bigtriangleup}{=}}
\newcommand{\oabf}{{\bf \breve{a}}}
\newcommand{\odbf}{{\bf \breve{d}}}
\newcommand{\oDbf}{{\bf \breve{D}}}
\newcommand{\oAbf}{{\bf \breve{A}}}
\newcommand{\calZbf}{\mbox{\boldmath $\cal Z$}}
\newcommand{\feop}{\hfill \rule{2mm}{2mm} \\}

\newcommand{\Rnum}{{\mathbb R}}
\newcommand{\Cnum}{{\mathbb C}}
\newcommand{\Znum}{{\mathbb Z}}
\newcommand{\Rset}{{\mathbb R}}
\newcommand{\Cset}{{\mathbb C}}
\newcommand{\Zset}{{\mathbb Z}}

\newcommand{\Ccal}{{\cal C}}
\newcommand{\Gcal}{{\cal G}}
\newcommand{\Rcal}{{\cal R}}
\newcommand{\Zcal}{{\cal Z}}
\newcommand{\Xcal}{{\cal X}}
\newcommand{\zzbf}{{\bf 0}}

\newcommand{\bma}{\mbox{\boldmath$a$}}
\newcommand{\bmb}{\mbox{\boldmath$b$}}
\newcommand{\bmc}{\mbox{\boldmath$c$}}
\newcommand{\bmd}{\mbox{\boldmath$d$}}
\newcommand{\bme}{\mbox{\boldmath$e$}}
\newcommand{\bmf}{\mbox{\boldmath$f$}}
\newcommand{\bmg}{\mbox{\boldmath$g$}}
\newcommand{\bmh}{\mbox{\boldmath$h$}}
\newcommand{\bmi}{\mbox{\boldmath$i$}}
\newcommand{\bmj}{\mbox{\boldmath$j$}}
\newcommand{\bmk}{\mbox{\boldmath$k$}}
\newcommand{\bml}{\mbox{\boldmath$l$}}
\newcommand{\bmm}{\mbox{\boldmath$m$}}
\newcommand{\bmn}{\mbox{\boldmath$n$}}
\newcommand{\bmo}{\mbox{\boldmath$o$}}
\newcommand{\bmp}{\mbox{\boldmath$p$}}
\newcommand{\bmq}{\mbox{\boldmath$q$}}
\newcommand{\bmr}{\mbox{\boldmath$r$}}
\newcommand{\bms}{\mbox{\boldmath$s$}}
\newcommand{\bmt}{\mbox{\boldmath$t$}}
\newcommand{\bmu}{\mbox{\boldmath$u$}}
\newcommand{\bmv}{\mbox{\boldmath$v$}}
\newcommand{\bmw}{\mbox{\boldmath$w$}}
\newcommand{\bmx}{\mbox{\boldmath$x$}}
\newcommand{\bmy}{\mbox{\boldmath$y$}}
\newcommand{\bmz}{\mbox{\boldmath$z$}}

\newcommand{\bmA}{\mbox{\boldmath$A$}}
\newcommand{\bmB}{\mbox{\boldmath$B$}}
\newcommand{\bmC}{\mbox{\boldmath$C$}}
\newcommand{\bmD}{\mbox{\boldmath$D$}}
\newcommand{\bmE}{\mbox{\boldmath$E$}}
\newcommand{\bmF}{\mbox{\boldmath$F$}}
\newcommand{\bmG}{\mbox{\boldmath$G$}}
\newcommand{\bmH}{\mbox{\boldmath$H$}}
\newcommand{\bmI}{\mbox{\boldmath$I$}}
\newcommand{\bmJ}{\mbox{\boldmath$J$}}
\newcommand{\bmK}{\mbox{\boldmath$K$}}
\newcommand{\bmL}{\mbox{\boldmath$L$}}
\newcommand{\bmM}{\mbox{\boldmath$M$}}
\newcommand{\bmN}{\mbox{\boldmath$N$}}
\newcommand{\bmO}{\mbox{\boldmath$O$}}
\newcommand{\bmP}{\mbox{\boldmath$P$}}
\newcommand{\bmQ}{\mbox{\boldmath$Q$}}
\newcommand{\bmR}{\mbox{\boldmath$R$}}
\newcommand{\bmS}{\mbox{\boldmath$S$}}
\newcommand{\bmT}{\mbox{\boldmath$T$}}
\newcommand{\bmU}{\mbox{\boldmath$U$}}
\newcommand{\bmV}{\mbox{\boldmath$V$}}
\newcommand{\bmW}{\mbox{\boldmath$W$}}
\newcommand{\bmX}{\mbox{\boldmath$X$}}
\newcommand{\bmY}{\mbox{\boldmath$Y$}}
\newcommand{\bmZ}{\mbox{\boldmath$Z$}}

\newcommand{\eop}{\hfill $\Box$}

\newcommand{\gs}{\mathop{\gtrless}\limits}

\newcommand{\circlambda}{\mbox{$\Lambda$
             \kern-.85em\raise1.5ex
             \hbox{$\scriptstyle{\circ}$}}\,}

\newcommand{\tr}{\mathop{\rm tr}}
\newcommand{\var}{\mathop{\rm var}}
\newcommand{\cov}{\mathop{\rm cov}}
\newcommand{\diag}{\mathop{\rm diag}}

\def\submbox#1{_{\mbox{\footnotesize #1}}}
\def\supmbox#1{^{\mbox{\footnotesize #1}}}

%
\newtheorem{Theorem}{Theorem}
\newtheorem{Definition}{Definition}
\newtheorem{Proposition}{Proposition}
\newtheorem{Lemma}{Lemma}
\newtheorem{Corollary}{Corollary}
\newtheorem{Example}{Example}
%
%
\newcommand{\ThmRef}[1]{\ref{thm:#1}}
\newcommand{\ThmLabel}[1]{\label{thm:#1}}
\newcommand{\DefRef}[1]{\ref{def:#1}}
\newcommand{\DefLabel}[1]{\label{def:#1}}
\newcommand{\PropRef}[1]{\ref{prop:#1}}
\newcommand{\PropLabel}[1]{\label{prop:#1}}
\newcommand{\LemRef}[1]{\ref{lem:#1}}
\newcommand{\LemLabel}[1]{\label{lem:#1}}
%


%% file: utils/comm.tex

\renewcommand \thetabf{\boldsymbol{\theta}}
\renewcommand \Phibf{\boldsymbol{\Phi}}
\renewcommand \Psibf{\boldsymbol{\Psi}}
\renewcommand \alphabf{\boldsymbol{\alpha}}
\renewcommand \betabf{\boldsymbol{\beta}}
\renewcommand \gammabf{\boldsymbol{\gamma}}
\renewcommand \deltabf{\boldsymbol{\delta}}
\renewcommand \epsilonbf{\boldsymbol{\epsilon}}
\renewcommand \zetabf{\boldsymbol{\zeta}}
\renewcommand \etabf{\boldsymbol{\eta}}
\renewcommand \iotabf{\boldsymbol{\iota}}
\renewcommand \kappabf{\boldsymbol{\kappa}}
\renewcommand \lambdabf{\boldsymbol{\lambda}}
\renewcommand \mubf{\boldsymbol{\mu}}
\renewcommand \nubf{\boldsymbol{\nu}}
\renewcommand \xibf{\boldsymbol{\xi}}
\renewcommand \pibf{\boldsymbol{\pi}}
\renewcommand \rhobf{\boldsymbol{\rho}}
\renewcommand \sigmabf{\boldsymbol{\signma}}
\renewcommand \taubf{\boldsymbol{\tau}}
\renewcommand \upsilonbf{\boldsymbol{\upsilon}}
\renewcommand \phibf{\boldsymbol{\phi}}
\renewcommand \varphibf{\boldsymbol{\varphi}}
\renewcommand \chibf{\boldsymbol{\chi}}
\renewcommand \psibf{\boldsymbol{\psi}}
\renewcommand \omegabf{\boldsymbol{\omega}}
\renewcommand \Sigmabf{\boldsymbol{\Sigma}}
\renewcommand \Upsilonbf{\boldsymbol{\Upsilon}}
\renewcommand \Omegabf{\boldsymbol{\Omega}}
\renewcommand \Deltabf{\boldsymbol{\Delta}}
\renewcommand \Gammabf{\boldsymbol{\Gamma}}
\renewcommand \Thetabf{\boldsymbol{\Theta}}
\renewcommand \Lambdabf{\boldsymbol{\Lambda}}
\renewcommand \Xibf{\boldsymbol{\Xi}}
\renewcommand \Pibf{\boldsymbol{\Pi}}

\renewcommand{\bma}{\boldsymbol{a}}
\renewcommand{\bmb}{\boldsymbol{b}}
\renewcommand{\bmc}{\boldsymbol{c}}
\renewcommand{\bmd}{\boldsymbol{d}}
\renewcommand{\bme}{\boldsymbol{e}}
\renewcommand{\bmf}{\boldsymbol{f}}
\renewcommand{\bmg}{\boldsymbol{g}}
\renewcommand{\bmh}{\boldsymbol{h}}
\renewcommand{\bmi}{\boldsymbol{i}}
\renewcommand{\bmj}{\boldsymbol{j}}
\renewcommand{\bmk}{\boldsymbol{k}}
\renewcommand{\bml}{\boldsymbol{l}}
\renewcommand{\bmm}{\boldsymbol{m}}
\renewcommand{\bmn}{\boldsymbol{n}}
\renewcommand{\bmo}{\boldsymbol{o}}
\renewcommand{\bmp}{\boldsymbol{p}}
\renewcommand{\bmq}{\boldsymbol{q}}
\renewcommand{\bmr}{\boldsymbol{r}}
\renewcommand{\bms}{\boldsymbol{s}}
\renewcommand{\bmt}{\boldsymbol{t}}
\renewcommand{\bmu}{\boldsymbol{u}}
\renewcommand{\bmv}{\boldsymbol{v}}
\renewcommand{\bmw}{\boldsymbol{w}}
\renewcommand{\bmx}{\boldsymbol{x}}
\renewcommand{\bmy}{\boldsymbol{y}}
\renewcommand{\bmz}{\boldsymbol{z}}

\renewcommand{\bmA}{\boldsymbol{A}}
\renewcommand{\bmB}{\boldsymbol{B}}
\renewcommand{\bmC}{\boldsymbol{C}}
\renewcommand{\bmD}{\boldsymbol{D}}
\renewcommand{\bmE}{\boldsymbol{E}}
\renewcommand{\bmF}{\boldsymbol{F}}
\renewcommand{\bmG}{\boldsymbol{G}}
\renewcommand{\bmH}{\boldsymbol{H}}
\renewcommand{\bmI}{\boldsymbol{I}}
\renewcommand{\bmJ}{\boldsymbol{J}}
\renewcommand{\bmK}{\boldsymbol{K}}
\renewcommand{\bmL}{\boldsymbol{L}}
\renewcommand{\bmM}{\boldsymbol{M}}
\renewcommand{\bmN}{\boldsymbol{N}}
\renewcommand{\bmO}{\boldsymbol{O}}
\renewcommand{\bmP}{\boldsymbol{P}}
\renewcommand{\bmQ}{\boldsymbol{Q}}
\renewcommand{\bmR}{\boldsymbol{R}}
\renewcommand{\bmS}{\boldsymbol{S}}
\renewcommand{\bmT}{\boldsymbol{T}}
\renewcommand{\bmU}{\boldsymbol{U}}
\renewcommand{\bmV}{\boldsymbol{V}}
\renewcommand{\bmW}{\boldsymbol{W}}
\renewcommand{\bmX}{\boldsymbol{X}}
\renewcommand{\bmY}{\boldsymbol{Y}}
\renewcommand{\bmZ}{\boldsymbol{Z}}